\documentclass[twocolumn,notitlepage,prx,superscriptaddress,longbibliography]{revtex4-1}
\usepackage{textcomp}
\usepackage{times}
\usepackage{graphicx}
\usepackage{float}
\usepackage{latexsym,amsmath,amssymb,bm,euscript}
\usepackage{color}
\usepackage{subfigure}
\usepackage{epstopdf}
\usepackage[colorlinks=true,linkcolor=blue,citecolor=blue]{hyperref}
\usepackage{hyperref}
\usepackage{soul}
\usepackage[normalem]{ulem}
\usepackage{mathrsfs}
\usepackage{amsmath}
\usepackage{xspace}
\usepackage{textcomp}

\newcommand{\Sec}[1]{Sec.~\ref{#1}}
\newcommand{\App}[1]{App.~\ref{#1}}
\newcommand{\Eq}[1]{Eq.~\eqref{#1}}

\newcommand{\Fig}[1]{Fig.~\ref{#1}}
\newcommand{\Figs}[1]{Figs.~\ref{#1}}

\def\Dstar{\ensuremath{D^\ast}\xspace}

\newcommand{\OS}[2]{OS\hspace{0.01in}$#1{\times}#2$}
\newcommand{\YC}[2]{YC\hspace{0.01in}$#1{\times}#2$}

\def\ucenter{\ensuremath{u_\mathrm{center}}\xspace} 
\def\utorus{\ensuremath{u_\mathrm{torus}}\xspace} 
\def\utot{\ensuremath{u_\mathrm{tot}}\xspace}
\def\usubtr{\ensuremath{u_\mathrm{subtr}}\xspace}

\def\TS{\ensuremath{T_S}\xspace}  
\def\mS{\ensuremath{m_S}\xspace}  

\graphicspath{{./Figs/}}

\begin{document}
\title{Thermal Tensor Renormalization Group Simulations of Square-Lattice Quantum Spin Models}

\author{Han Li}
\affiliation{Department of Physics, Key Laboratory of Micro-Nano Measurement-Manipulation and Physics (Ministry of Education), Beihang University, Beijing 100191, China}

\author{Bin-Bin Chen}
\affiliation{Department of Physics, Key Laboratory of Micro-Nano Measurement-Manipulation and Physics (Ministry of Education), Beihang University, Beijing 100191, China}
\affiliation{Munich Center for Quantum Science and Technology (MCQST), Arnold Sommerfeld Center for Theoretical Physics (ASC) and Center for NanoScience (CeNS), Ludwig-Maximilians-Universit\"at M\"unchen, Fakult\"at f\"ur Physik, D-80333 M\"unchen, Germany}

\author{Ziyu Chen}
\affiliation{Department of Physics, Key Laboratory of Micro-Nano Measurement-Manipulation and Physics (Ministry of Education), Beihang University, Beijing 100191, China}

\author{Jan von Delft}
\affiliation{Munich Center for Quantum Science and Technology (MCQST), Arnold Sommerfeld Center for Theoretical Physics (ASC) and Center for NanoScience (CeNS), Ludwig-Maximilians-Universit\"at M\"unchen, Fakult\"at f\"ur Physik, D-80333 M\"unchen, Germany}

\author{Andreas Weichselbaum}
\email{weichselbaum@bnl.gov}
\affiliation{Department of Condensed Matter Physics and Materials
Science, Brookhaven National Laboratory, Upton, New York 11973-5000, USA}
\affiliation{Munich Center for Quantum Science and Technology (MCQST), Arnold Sommerfeld Center for Theoretical Physics (ASC) and Center for NanoScience (CeNS), Ludwig-Maximilians-Universit\"at M\"unchen, Fakult\"at f\"ur Physik, D-80333 M\"unchen, Germany}

\author{Wei Li}
\email{w.li@buaa.edu.cn}
\affiliation{Department of Physics, Key Laboratory of Micro-Nano Measurement-Manipulation and Physics (Ministry of Education), Beihang University, Beijing 100191, China}
\affiliation{International Research Institute of Multidisciplinary Science, Beihang University, Beijing 100191, China}

\begin{abstract}
In this work, we benchmark the well-controlled and numerically
accurate exponential thermal tensor renormalization group (XTRG)
in the simulation of interacting spin models in two dimensions.
Finite temperature introduces a thermal correlation length,
which justifies the analysis of finite system size for the sake
of numerical efficiency. In this paper we focus on the square
lattice Heisenberg antiferromagnet (SLH) and quantum Ising
models (QIM) on open and cylindrical geometries up to width
$W=10$. We explore various one-dimensional mapping paths in the
matrix product operator (MPO) representation, whose performance
is clearly shown to be geometry dependent. We benchmark against
quantum Monte Carlo (QMC) data, yet also the series-expansion
thermal tensor network results. Thermal properties including
the internal energy, specific heat, and spin structure factors,
etc., are computed with high precision, obtaining excellent
agreement with QMC results. XTRG also allows us to reach
remarkably low temperatures. For SLH we obtain at low
temperature an energy per site $u_g^*\simeq -0.6694(4)$ and a
spontaneous magnetization $\mS^*\simeq0.30(1)$, which is
already consistent with the ground state properties. We extract an
exponential divergence vs. $T$ of the structure factor $S(M)$,
as well as the correlation length $\xi$, at the ordering wave
vector $M=(\pi,\pi)$, which represents the renormalized
classical behavior and can be observed over a narrow but
appreciable temperature window, by analysing the finite-size
data by XTRG simulations. For the QIM with a
finite-temperature phase transition, we employ several thermal
quantities, including the specific heat, Binder ratio, as well
as the MPO entanglement to determine the critical temperature
$T_c$.
\end{abstract}

\date{\today} 
\maketitle

\section{Introduction}

Two-dimensional (2D) lattice models play an important role in
our understanding of correlated quantum materials
\cite{Chakravarty1989,Greven94,Elstner95,Dagotto05,Rawl2017}.
Their efficient simulation, however, constitutes a major
challenge in contemporary condensed matter physics and beyond.
Renormalization group (RG) methods, including the density matrix
renormalization group (DMRG) \cite{PhysRevLett.69.2863} and
other tensor-network based RG algorithms
\cite{Verstraete.f+:2004:2DRenormalization, Verstraete2008} have
been established as powerful tools solving 2D many body problems
at $T=0$. They have achieved success in searching for quantum
spin liquids (QSLs) in 2D frustrated magnets, e.g., Kagome-
\cite{Yan2011,Depenbrock.s+:2012:SpinLiquid} and
triangular-lattice
\cite{PhysRevLett.99.127004,Zhu2015,Hu2015,PhysRevLett.120.207203}
Heisenberg models, etc.

Finite-temperature properties can also be simulated by RG-type
algorithms, e.g., the transfer-matrix renormalization group
(TMRG) \cite{Bursill.r.j+:1996:DMRG,Wang.x+:1997:TMRG,
Xiang.t:1998:Thermodynamics}. TMRG finds the dominating
eigenstate as well as corresponding eigenvalue of the transfer
matrix by using the DMRG algorithm, and thus obtains thermal
properties directly in the thermodynamic limit. Besides, for a
finite-size system, the finite-$T$ DMRG scheme
\cite{Feiguin.a.e+:2005:ftDMRG} using imaginary-time evolution,
and an algorithm based on the minimally entangled typical
thermal states \cite{White.s.r:2009:METTS,
Stoudemire.e.m+:2010:METTS}, have been proposed. Although the
above thermal RG methods are very successful in one dimension
(1D), their efficient generalization to 2D constitutes a very
challenging task.

Among others, the linearized tensor renormalization group (LTRG)
approach contracts the thermal tensor network (TTN) linearly in
the ``imaginary time", i.e., inverse temperature $\beta$
\cite{Li.w+:2011:LTRG}, typically in a Trotterized scheme, and
can be employed to simulate infinite- and finite-size 1D systems
\cite{Dong.y+:2017:BiLTRG}. By expressing corresponding thermal
states as tensor product operators (TPO), LTRG can be employed
to simulate infinite 2D lattices
\cite{Ran.s+:2012:Super-orthogonalization,
Czarnik.p+:2012:PEPS,Czarnik.p+:2015:PEPS,Czarnik.p+:2017:Sign,
Czarnik.p+:2016:TNR}. However, due to the approximations as
well as large computational costs in the tensor optimization
scheme, precise and highly controllable TPO methods are still
under exploration.
 
On the other hand, TTN methods for finite-size 2D systems have
been put forward only recently, using matrix product operator
(MPO) representations of the density matrix
\cite{Bruognolo.b+:2017:MPS, Chen.b+:2017:SETTN, Chen2018}.
These MPO-based approaches, in particular, series-expansion TTN
(SETTN) \cite{Chen.b+:2017:SETTN} and exponential tensor
renormalization group (XTRG) \cite{Chen2018}, are controlled,
quasi-exact methods that are highly competitive when tackling
even very challenging problems in 2D \cite{Chen2018a}.

In this work, we explore the square lattice Heisenberg (SLH) and
the quantum Ising model (QIM) under transverse fields, with the
above-mentioned MPO thermal RG methods, aiming to benchmark the
accuracy. The obtained thermal data are compared to quantum
Monte Carlo (QMC) results, where excellent agreement is
observed. We perform a thorough (truncation) error and
finite-size analysis which allows us to extract low-energy down
to ground-state properties including ground state energy and
spontaneous magnetization. Similarly, we analyze the critical
temperature of thermal phase transition, etc., and compare all
of these to well established QMC results.

The rest of the paper is organized as follows.
Sec.~\ref{Sec:MoMe} introduces the spin lattice models and the
TTN methods, as well as thermal quantities concerned in
the present work. In Sec.~\ref{Sec:Geo}, we compare four
different MPO mapping paths (see \Fig{Fig:maxPe} below), and
find the \textit{snake}-like path, usually employed in ground
state computations, also to be the overall most efficient one in
our thermal simulations. Our main results for the SLH and QIM
are discussed in Sec. \ref{Sec:SLH} and Sec. \ref{App:QIM}. The
last section is devoted to a summary. 

\section{Models and Methods}
\label{Sec:MoMe}

\subsection{Quantum Spin Models on the Square Lattice}
\label{Sec:Models}

A paradigmatic model in quantum magnetism 
is the square lattice Heisenberg (SLH) 
antiferromagnet whose Hamiltonian reads
\begin{equation}
H = J \sum_{\langle i,j \rangle} {S}_i \, \cdot \, {S}_j,
\label{Eq:SLH}
\end{equation}
where $J$ is the coupling strength of isotropic spin
interactions between nearest-neighbors (NN), as denoted by
$\langle .,. \rangle$. The SLH is a simple yet fundamental
quantum lattice model of interacting spins, and hence of great
interest on its own. It can be derived as the large $U$ limit of
the Hubbard model at half-filling
\cite{Hubbard238,*Hubbard1979a,*Hubbard1979b,*Hubbardin1981}, 

There exists true long-range N\'eel order in the ground state of
SLH \cite{Anderson1952order, ordercalcu1, ordercalcu2,
ordercalcu3} which, nevertheless, according to the renowned
Mermin-Wagner theorem \cite{Mermin-Wagner}, ``melts" immediately
when thermal fluctuations are introduced. However, incipient
order formed by correlated large-size clusters is still present
at low temperatures, i.e., in the so-called renormalized
classical (RC) regime, where the sizes of ordered clusters,
i.e., the correlation length $\xi$, increase exponentially as
temperature is lowered \cite{Chakravarty1989, Manousakis1991}.

Besides SLH, we also apply our thermal RG methods to study the
quantum Ising model (QIM), 
\begin{equation}
H = - J \sum_{\langle i,j \rangle} S^z_i S^z_j + h \sum_i S^x_i,
\label{Eq:QIM}
\end{equation}
again with NN coupling $J$, 
$S^{x(z)}$ is the $x(z)$ component of the spin operator, 
and $h$ is the transverse field. At $T=0$, 
a quantum phase transition (QPT) takes place at 
$h_c \simeq 1.52219(1)$ \cite{TFI2DQMC}:
for $h< h_c$ the system is ferromagnetically (FM) ordered, 
while for $h> h_c$ it is in a quantum paramagnetic phase. 
In the former case, thermal fluctuations drive 
a phase transition at $T=T_c$, 
above which the system enters a classical paramagnetic phase. 
The determination of critical temperature 
$T_c$ constitutes another interesting benchmark for XTRG.

\begin{figure}[!tbp]
\includegraphics[angle=0,width=0.84\linewidth]{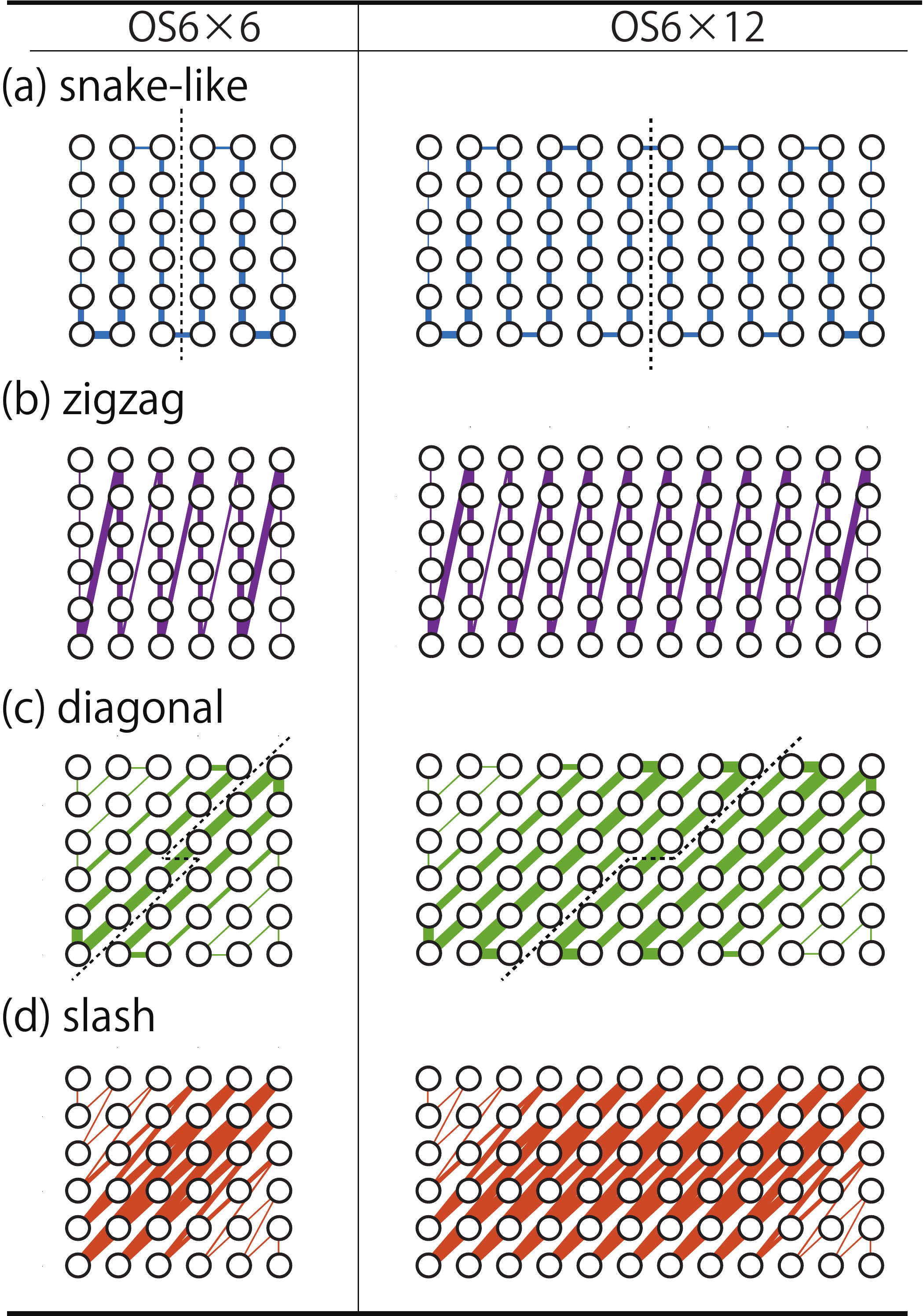}
\caption{(Color online) 
   Various MPO paths utilized in XTRG simulations that map the 2D
   lattices into quasi-1D system with long-range interactions,
   including (a) the \textit{snake}-like, (b) \textit{zigzag}, (c)
   \textit{diagonal}, and (d) \textit{slash} paths. The line width
   visualizes the low-temperature bond entanglement $S_{E}$ along
   the MPO obtained on the \OS{6}{6} and \OS{6}{12} lattices (at
   $T\simeq0.06$), where we used a width of $w = (4 S_E - 11)
   \,\mathrm{pts}$, yet enforcing $w\geq1$ for visibility. 
}
\label{Fig:maxPe}
\end{figure}

In our simulations below, we mainly consider two different
square-lattice geometries. These are the open strip (OS) $W
\times L$ geometries for system with width $W$ and length $L$,
and cylindrical lattice (YC) $W\times L$ systems wrapped along
the width $W$ in the vertical y-direction w.r.t. the MPO paths
shown in \Fig{Fig:maxPe}. Throughout this paper we use $J=1$ as
the unit of energy, lattice spacing $a=1$, and Boltzmann
constant $k_B=1$.

\subsection{Thermal Tensor Renormalization Group Methods}
\label{Sec:Methods}

We employ thermal tensor renormalization group (TRG) methods,
including XTRG and SETTN, to simulate the spin lattice models.
In both approaches, the {\it unnormalized} density matrix
$\rho(\beta)\equiv e^{-\beta H}$ of a finite-size 2D system is
represented in terms of MPO in a quasi-1D setup. In XTRG,
$\rho(\tau)$ at small inverse temperature $\tau$ is initialized
through a Taylor expansion, i.e., \begin{equation} \rho(\tau)
\simeq \sum_{k=0}^{N_c} \tfrac{(-\tau)^k}{k!} H^k,
\label{Eq:Mac} \end{equation} with $N_c$ the cut-off order. The
RG techniques required to construct efficient TTN representations
of the initial $\rho(\tau)$ have been developed in the SETTN
algorithms \cite{Chen.b+:2017:SETTN}.

After the initialization, we double the inverse temperature
$\beta = 2^n \cdot \tau$ of the density matrix $\rho_n$ in each
iteration $n$ and thus cool down the system exponentially fast,
i.e.,
\begin{equation}
 \rho_{n+1} = \rho_n * \rho_n,
\label{Eq:XTRG}
\end{equation}
where $*$ indicates MPO multiplication. XTRG turns out to be
very efficient and accurate (compared to linearly decreasing the
temperature, it yields smaller accumulated truncation errors due
to significantly less truncation steps). It can be parallelized
via a $z$-shift of the initial $\tau$, i.e., $\tau \rightarrow
2^z \tau$ with $z \in [0,1)$, to obtain fine-grained temperature
resolution \cite{Chen2018}. Overall, our approach is equivalent
to the purification framework \cite{Feiguin.a.e+:2005:ftDMRG,
Barthel2009, Li.w+:2011:LTRG, Dong.y+:2017:BiLTRG,
Zwolak.m+:2004:Superoperator}, and $T_\mathrm{min}\equiv
1/\beta_\mathrm{max}$ constitutes the lowest temperature
reached.

Apart from providing a good initialization for small $\tau$,
SETTN also provides an alternative way to determine
$\rho(\beta)$ for simulations down to low temperatures, also
operating on a logarithmic $\beta$ grid. To be specific, a
point-wise Taylor expansion version of SETTN, proposed in
Ref.~\cite{Chen2018}, is adopted in this work. It expands the
thermal state 
\begin{equation}
  \rho(\beta) = 
  \sum_{k=0}^{N_c} \tfrac{(-\beta+\beta_n)^k}{k!} H^k \rho(\beta_n),
\label{Eq:Tay}
\end{equation} 
around a series of temperature points $\beta_n \to 2\beta_n$
starting at $\beta_0=\tau$, such that $\beta\in \{2\tau, 4\tau,
\ldots, 2^n \tau \equiv \beta_{\mathrm{max}}/2 \}$ for XTRG,
as well as smaller $\beta$ steps in case of SETTN. Since
truncation errors accumulate as $k$ increases in each $H^k
\rho(\beta_n)$ term of the series, this modified SETTN reduces
the order $N_c$ required for the expansion thus improves the
accuracy. Besides, the SETTN approach also benefits in
efficiency from the logarithmic scales in temperature series
$\{\beta_n\}$, since it reduces significantly the computational
overhead in expansions.

\begin{figure}[!tbp]
\includegraphics[angle=0,width=1\linewidth]{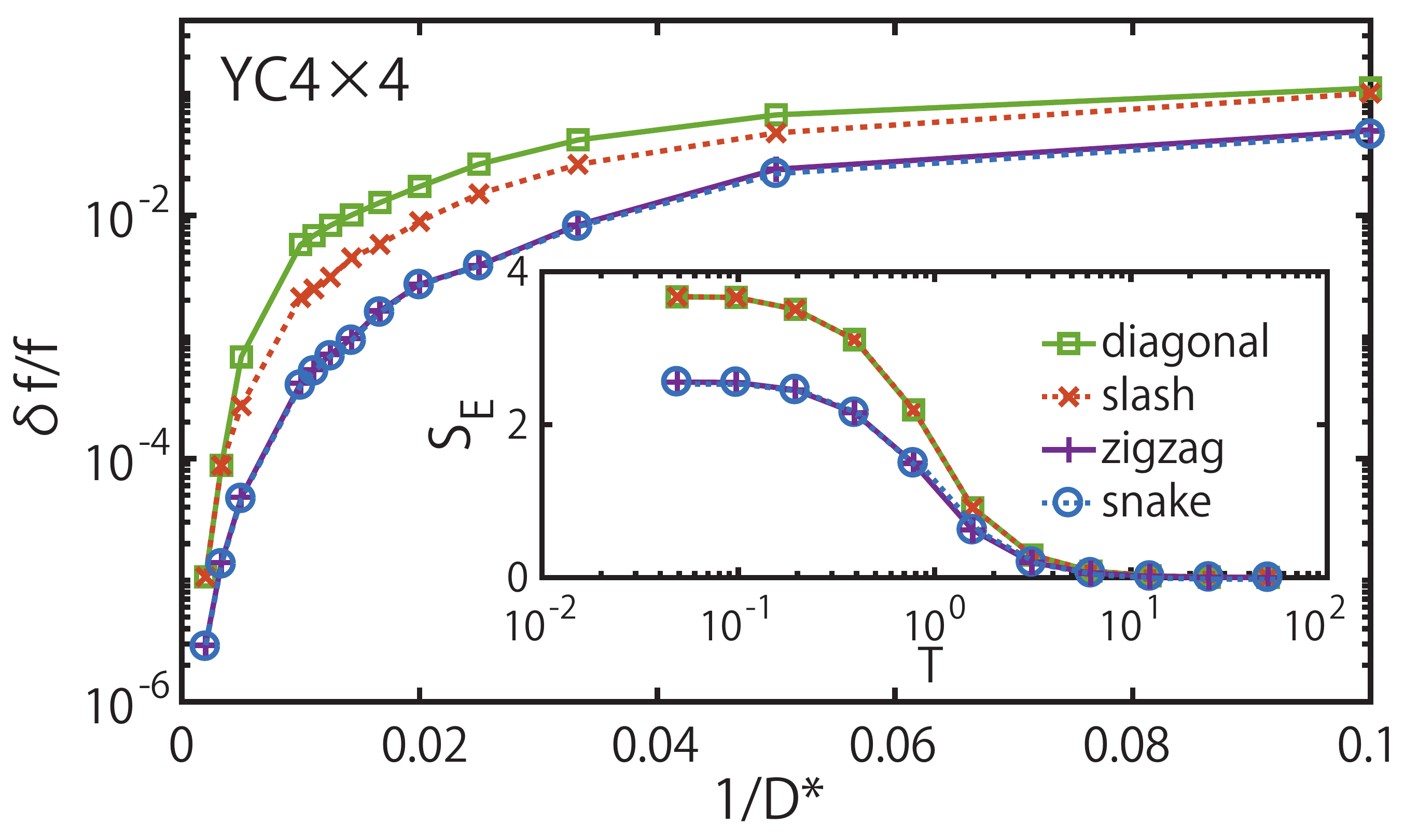}
\caption{(Color online) 
   Relative errors of free energy $f$ vs. $1/D^*$, $D^*$ the number
   of multiplets kept, for the \YC{4}{4} SLH at $T\simeq0.06$, and
   the ED data are taken as the exact reference. Four mapping paths
   are compared, including the \textit{snake}-like, \textit{slash},
   \textit{diagonal}, and the \textit{zigzag} one. The inset shows
   the maximum of $S_E$ over all MPO bonds, vs. $T$. One can
   observe that $S_E$ values coincide in the essentially equivalent
   \textit{snake}-like and \textit{zigzag} paths, and are
   significantly smaller compared to the \textit{diagonal} and
   \textit{slash} paths. 
}
\label{Fig:Fe4x4}
\end{figure}

\subsection{Thermal quantities and entanglement measurements}
\label{Sec:entanglement}
In this work, we are interested in various quantities, 
including the free energy $f$, internal energy $u$, specific heat $c_V$, 
and static magnetic structure factor $S({q})$,
as well as MPO entanglement $S_E$ in the thermal states.

The free energy per site can be directly computed
from the partition function, 
\begin{equation}
f(\beta)=-\tfrac{1}{\beta N} \ln{\mathcal{Z}(\beta)},
\end{equation} 
where $\mathcal{Z}(\beta) = \rm{Tr}\bigl[\rho(\frac{\beta}{2})
^\dagger \,  \rho(\frac{\beta}{2})\bigr]$ is the partition function
and $N$ is the total number of sites. The internal energy $u$
per site can be evaluated, in practice, in two different yet
theoretically equivalent ways. A simple way is to compute the
expectation value $u(\beta)$ directly by tracing the total
Hamiltonian $H$ with density operators $\rho$ (referred to as
scheme $a$), 
\begin{subequations} \label{eq:udef} 
\begin{equation}
   u(\beta) = \tfrac{1}{N} \mathrm{Tr} \bigl[ 
      \rho(\tfrac{\beta}{2})^{\dagger}  H  \rho(\tfrac{\beta}{2})
   \bigr].
\label{Eq:uTr}
\end{equation}
Since the MPO representations of the density matrices $\rho$ and
Hamiltonian $H$ are available in XTRG and SETTN simulations,
Eq.~(\ref{Eq:uTr}) can be calculated conveniently via tensor
contractions. Alternatively, one can also compute the internal
energy $u$ by taking derivatives of free energy $f$ (referred to
as scheme $b$), 
\begin{equation}
u(\beta) = \tfrac{1}{N} \tfrac{\partial(f\beta)}{\partial \beta} = \tfrac{1}{N \beta} \tfrac{\partial (f\beta)}{\partial \ln{\beta}},
\label{Eq:uDr}
\end{equation}
\end{subequations} 
where the last derivative is a natural choice when $\beta$ is
chosen on a logarithmic grid. The specific heat $c_V$ is given
by the derivative of the internal energy, 
\begin{equation}
   c_V = \tfrac{\partial u}{\partial T} 
       = - {\beta} \tfrac{\partial u}{\partial \ln{\beta}},
\label{Eq:Cv}
\end{equation}
again with preference to taking the derivative with respect to
the logarithmic temperature scale, as shown in the last term.

In order to understand the spin structure at finite $T$, e.g.,
to probe the incipient order and estimate the spontaneous
magnetization in the SLH model, we compute the static spin
structure factor $S({q})$ at finite temperature, defined as 
\begin{equation}
   S(q) = \sum_{j} e^{-i q\cdot r_{ij}} \langle S_{i} \cdot S_{j} \rangle_T, 
\label{Eq:SQ}
\end{equation}
where $r_{ij} \equiv r_j-r_i$ refers to the distance between
lattice site $i$ and $j$. Dealing with finite system sizes, we
fix $i$ in the center of the system, whereas $j$ runs over the
entire lattice. 

By choosing $q$ in the vicinity of the ordering wave vector $q_0
=M\equiv (\pi, \pi)$ [cf.~\Fig{Fig:u2}(e)], one has $S(q)
= S(q_0)/[1+\xi^2(q-q_0)^2]$ (Ornstein-Zernike form), and thus
$\xi^2 \cong \bigl. \tfrac{-1}{2S({q})} \tfrac{\partial^2
S({q})}{\partial {q}^2} \bigr|_{{q}={q}_0}$, from which it
follows \cite{Elstner-1993}
\begin{equation}
\xi^2(T) = \tfrac{c^2_{{q}_0}}{2S({q}_0)} \sum_j
  {r}_{ij}^2 \, e^{-\text{i} {q}_0 \cdot {r}_{ij}} \,
  \langle {S}_i \cdot {S}_j \rangle_T
 \text{ ,}\label{Eq:Xi}
\end{equation}
where the constant $c^2_{{q}_0} \equiv \langle \cos^2
\alpha_{ij}\rangle = 1/2$ accounts for an angular average, with
$\alpha_{ij}$ the angle in between ${q}_0$ and ${r}_{ij}$. 

We also investigate the MPO entanglement, which offers direct
information on the numerical efficiency of our thermal RG
simulations. In XTRG, the MPO density matrix can be regarded as
a purified \textit{superstate} $|\tilde{\Psi}_{\beta/2} \rangle
\hat{=} \rho(\beta/2)$, which is unnormalized, hence the tilde.
By definition then, the partition function can be calculated as
$\mathcal{Z}(\beta) = \langle \tilde{\Psi}_{\beta/2} |
\tilde{\Psi}_{\beta/2} \rangle$. This thermofield double
purification employs identical ancillary and physical state
spaces.
It is then useful to introduce a formal entanglement measure,
$S_E$, for the MPO. For this, we divide the {\it normalized}
super-vector $| \Psi_{\beta/2} \rangle$ (hence no tilde) --
represented now as an effective matrix product state (MPS) with
twice as many, paired up local degrees of freedom -- into two
blocks w.r.t. some specified bond, and compute the standard MPS
block-entanglement (von Neumann) entropy $S_E$
\cite{Schollwoeck11,Chen2018}. The latter is a measure of both
quantum entanglement and classical correlations. As such, $S_E$
is a quantity of practical importance in our thermal RG
simulations, since the bond dimension $D \sim e^{S_E}$
quantifies the required computational resource for an accurate
description of the thermal states.

In conformal quantum critical chains, the MPO entanglement $S_E$
scales logarithmically vs. $\beta$, as derived from conformal
field theory \cite{Dubail17} and confirmed in large-scale numerics
\cite{Prosen.t+:2007:Entropy,Barthel.t:2017:FiniteT,Chen2018}.
The temperature dependence of $S_E$ strongly depends on the
underlying physics. In the following, it will be analyzed in
detail in this regard for the SLH, which has low-energy gapless
modes due to the spontaneous SU(2) symmetry breaking, as well as
in the QIM, which undergoes a finite-$T$ phase transition.

In our XTRG simulations of the SLH, finally, we also fully
exploit the global SU(2) symmetry in the MPO based on the QSpace
tensor library \cite{Weichselbaum.a:2012:QSpace}. In these
SU(2) symmetric calculations, a state-based description of any
state space or index is replaced in favor of a description in
terms of multiplets. Specifically, $D$ states on the geometric
MPO bonds are equivalently reduced to $\Dstar \simeq D/4$
multiplets, with \Dstar the tuning parameter.
Given the numerical cost for XTRG being $\mathcal{O}(D^4)$
\cite{Chen2018}, the implementation of non-Abelian symmetry in
XTRG therefore greatly improves its computational efficiency.
Conversely, this allows us to reach lower temperatures.

\begin{figure}[!tbp]
\includegraphics[angle=0,width=1\linewidth]{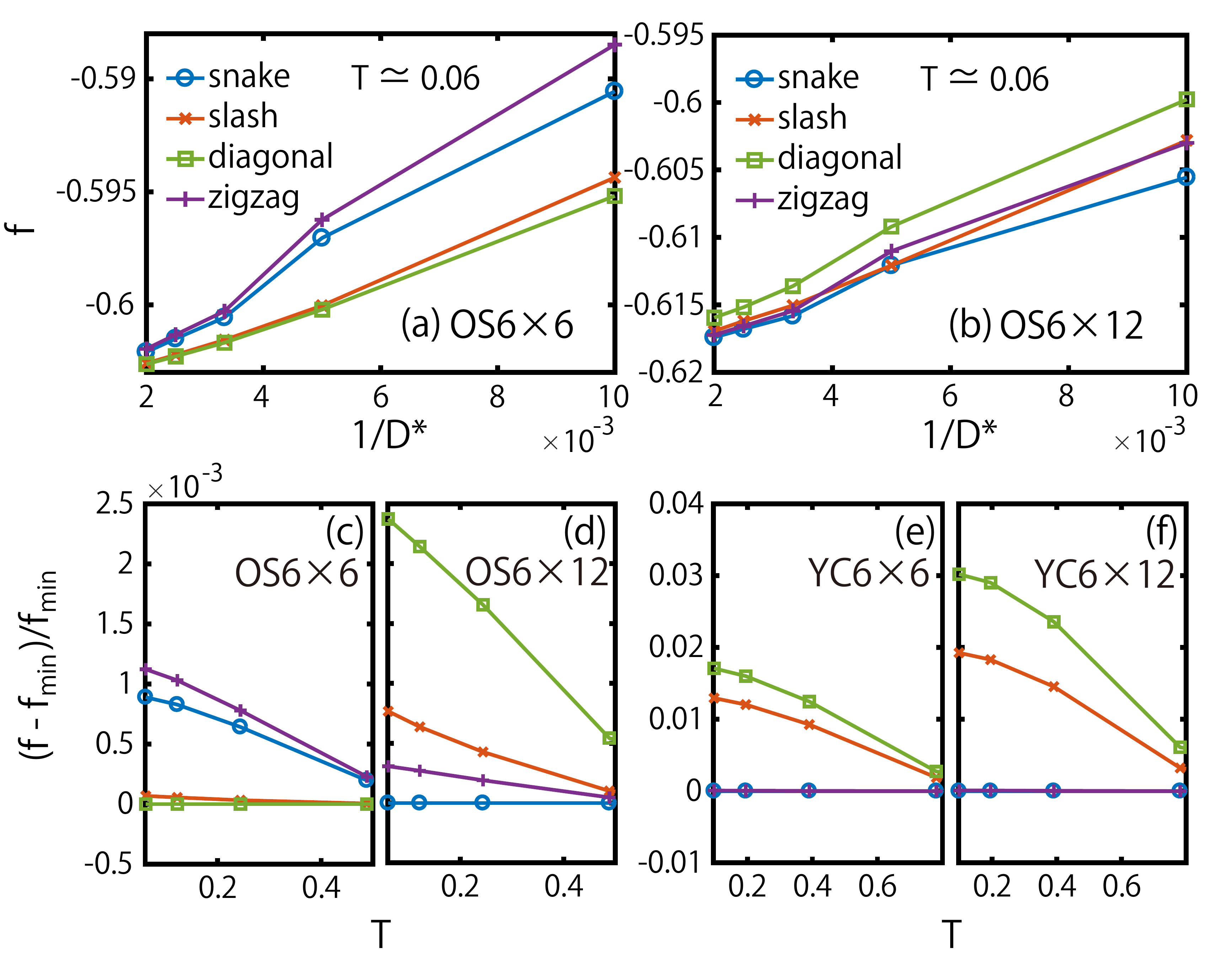}
\caption{(Color online)
   The free energy $f$ of SLH on the (a) \OS{6}{6} and (b)
   \OS{6}{12} lattices at $T\simeq0.06$, obtained for the four
   different MPO paths in \Fig{Fig:maxPe} by retaining $\Dstar=100$
   to $500$ multiplets in all cases. (c, d, e, f) show comparisons
   of the free energy $f$ vs. $T$ for all paths using $\Dstar=500$
   for OS (c,d) and YC (e,f). Here $f_{\mathrm{min}}(T)$
   represents the minimal value amongst all four paths at any
   given $T$. 
}
\label{Fig:FePath}
\end{figure}

\begin{figure*}[!tbp]
\includegraphics[angle=0,width=0.75\linewidth]{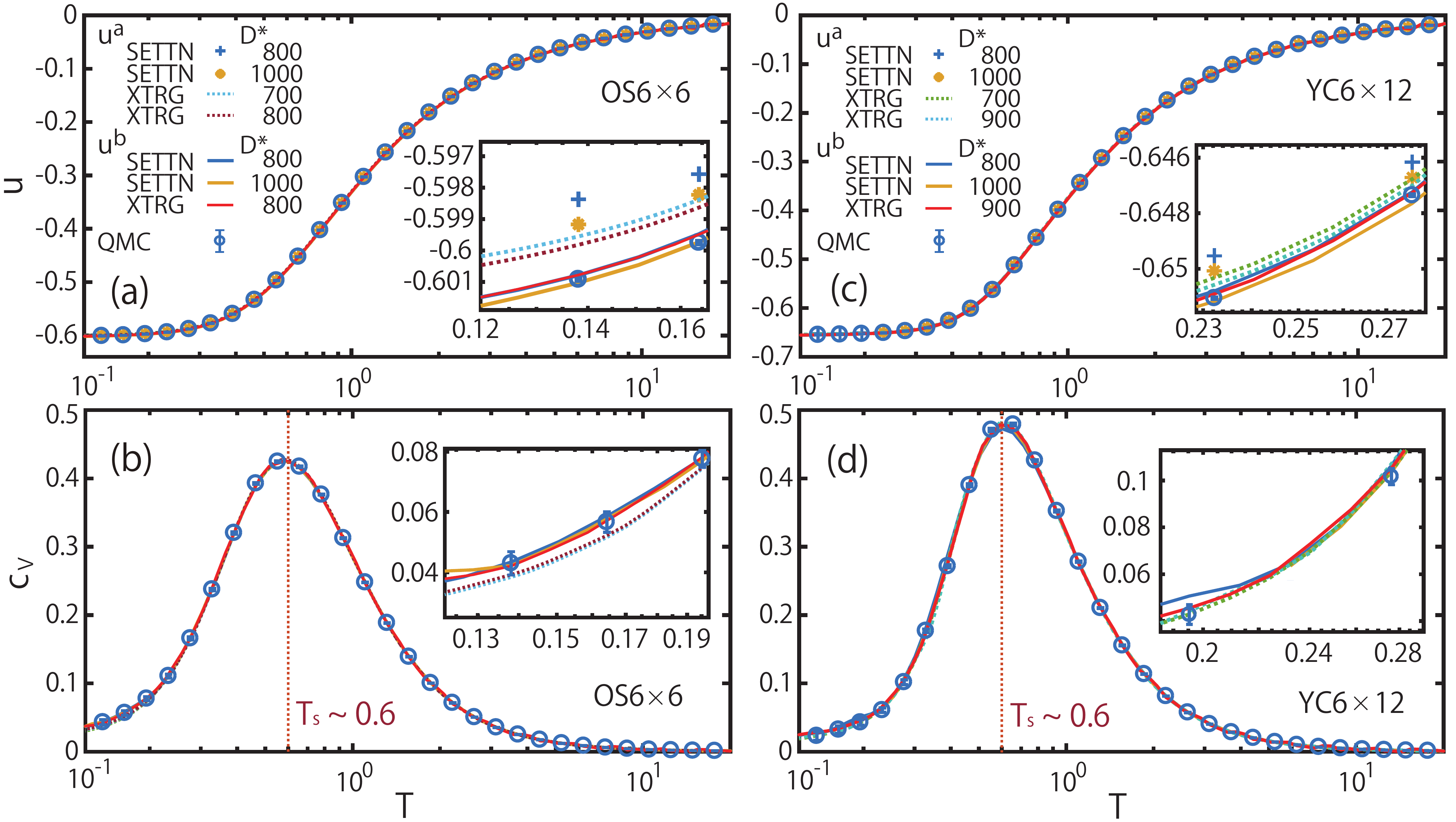}
\caption{(Color online)
   Internal energy $u$ and specific heat $c_V$ for (a,b) \OS{6}{6},
   and (c,d) \YC{6}{12} SLH systems. The insets zoom into the
   low-$T$ data, where the SETTN and XTRG data for various $D^*$
   are shown to agree excellently with QMC. 
   In the legends, $u^a$ and $u^b$ refer to the two schemes in
   Eqs.~\ref{eq:udef}\,(a) and (b), respectively. The specific
   heat $c_V$ in the lower panels is obtained from $u$ using
   \Eq{Eq:Cv}. The vertical dashed line represents the temperature
   scale $\TS \sim 0.6$ in SLH.
}
\label{Fig:W6uCv}
\end{figure*}

\section{Various MPO paths in thermal renormalization group simulations}
\label{Sec:Geo}

Since our MPO-based RG methods map the 2D lattice models into a
quasi-1D setup, the sites of the lattice must be brought into a
serial order. This introduces a `mapping path' throughout the
lattice, the specific choice of which clearly includes some
arbitrariness. 
This has already been discussed before in a similar context in
DMRG simulations \cite{Xiang2001}. There the authors considered
ordering the sites along the diagonal direction [cf.
Fig.~\ref{Fig:maxPe}(c)], made some comparisons to the
conventional \textit{snake}-like path [cf.
Fig.~\ref{Fig:maxPe}(a)], and arrived at a conclusion that the
\textit{diagonal} path gets better, i.e., lower variational
energy, when the same number of bond states is retained. Here
we perform a similar analysis for our thermal simulations. For
comparison, we include a few more conventional paths in our
thermal RG simulations, with the expectation to recover the
observations made in previous DMRG study mentioned above for the
same geometry.

To be specific, in \Fig{Fig:maxPe} we compare four simple
choices of paths: the \textit{snake}-like (blue color),
\textit{slash} (orange), \textit{diagonal} (green), and the
\textit{zigzag} one (purple). We perform XTRG calculations down
to low temperatures for these MPO paths on systems including $4
\times 4$ (YC), $6 \times 6$, and $6 \times 12$ (both OS and YC)
geometries. Throughout this section (as well as in
\App{App:MPOPath}), the same color code is adopted in all
related plots, e.g.,
Figs.~\ref{Fig:maxPe},\ref{Fig:Fe4x4},\ref{Fig:FePath} as well
as \Fig{Fig:truncPe}. 

Firstly, we benchmark the SLH on a small \YC{4}{4} lattice also
accessible by exact diagonalization (ED), by checking the
relative error of the free energy $f$ at a low temperature
($T\simeq0.06$) in Fig.~\ref{Fig:Fe4x4}. Clearly, $\delta f/f$
improves continuously with increasing $D^*$, down to $\sim
10^{-5}$ for $D^*=500$ retained bond multiplets. Overall, we
conclude from \Fig{Fig:Fe4x4}, that the \textit{snake}-like and
the \textit{zigzag} paths turn out to be optimal amongst all
four choices. 

However, the conclusion reached depends on the system size,
specifically so for smaller ones. In Figs.~\ref{Fig:FePath}(a,b),
we compare four MPO paths on the larger \OS{6}{6} and \OS{6}{12}
systems, where $f$ is compared at $T\simeq0.06$. Although ED
data is no longer available to compare to, the XTRG results for
$f$ are variational. Therefore a lower value of $f$ {\it still}
unambiguously serves as a useful criterion for accuracy.
In Figs.~\ref{Fig:FePath}(a,c), for the \OS{6}{6} system, we
find the \textit{diagonal}, as well as the \textit{slash} path,
leads to a lower, thus better, $f$, by a relative difference
$\lesssim 0.1\%$. This is in agreement with the observation in
Ref.~\cite{Xiang2001}, where they also find that the
\textit{diagonal} path produces energetically better results.

However, the situation quickly reverses again for larger
systems, and in particular also for the cylindrical geometries.
On the longer \OS{6}{12} lattice [Figs.~\ref{Fig:FePath}(b,d)],
the \textit{snake}-like path produces lower results for $f$,
closely followed by the zig-zag, while the \textit{diagonal} one
now leads to highest $f$ amongst all four choices, still with
relative differences $\lesssim 0.2\%$. For the YC geometries,
as shown in Figs.~\ref{Fig:FePath}(e,f), the \textit{snake}-like
path is again found to be the optimal choice, and the
\textit{diagnal} path the least favorable one, with $f$ now a
few percent larger at our lowest temperatures.
Conversely, the \textit{snake}-like and the \textit{zigzag}
paths show strong consistency within $10^{-4}$ relative
difference.

To shed some light on understanding the performance of various
mapping paths, we show the landscape of thermal entanglement
$S_E$ vs. MPO bond indices in Fig.~\ref{Fig:maxPe}, where the
bond thickness represents the ``strength" of entanglement. From
Fig.~\ref{Fig:maxPe}, as well as Fig.~\ref{Fig:truncPe}, one can
see that the \textit{snake}-like and the \textit{zigzag} path
have a comparatively small entanglement throughout their paths.
To be specific, for the \textit{snake}-like and \textit{zigzag}
paths, the bond entanglement distribution is rather uniform
(except for few bonds near both ends). By contrast, for the
\textit{slash} and \textit{diagonal} paths, there exist numerous
thick lines in the bulk, leading to overall larger truncation
errors (see \App{App:MPOPath} for detailed data)
and thus higher free energy results. 

One can understand the entanglement ``strength", as well as
required bond dimensions, on a given MPO bond in a somewhat
intuitive way: since we divide the system into two halves by
cutting only one MPO bond, it is natural to associate the
required bond dimension to the smallest possible number of
coupling bonds (lattice links) intersected by that specific
cutting line (see, e.g., dashed lines in Fig.~\ref{Fig:maxPe}).
For \OS{6}{6} (left column of Fig.~\ref{Fig:maxPe}), in the
\textit{snake}-like and \textit{zigzag} paths, the typical
bipartition line cuts 6 interaction links, while for the
\textit{diagonal} and \textit{slash} cases, this number is 10.
Note also that when the dashed cutting line has a corner, it can
introduce some additional constant contribution to the MPO
entanglement, which helps understand the specific location of
``thick" bonds in various paths in Fig.~\ref{Fig:maxPe}.
While for \OS{6}{6} one may argue, that entanglement only
concentrates on the narrow (anti-)diagonal and hence may be
beneficial, for more general geometries, say, long \OS{6}{12},
shown in the right column of Fig.~\ref{Fig:maxPe} (as well as in
cylindrical geometries, not shown), the \textit{snake}-like or
\textit{zigzag} path clearly constitutes a better choice.

To summarize, except for \OS{6}{6} where the
\textit{diagonal}-path has a slightly better performance,
indeed, in agreement with previous DMRG results
\cite{Xiang2001}, for larger systems the \textit{snake}-like or {\it
zigzag} paths are generally expected to lead to lower free
energy. Overall, we observe that from a computational and
accuracy point of view, \textit{zigzag} and \textit{snake}-like
paths are essentially equivalent and, in certain ways, so are
\textit{slash} to \textit{diagonal} paths. As expected and
shown explicitly in \Fig{Fig:FePath}, the accuracy for all paths
increases with increasing \Dstar. Nevertheless, this barely
changes the preference on a given path. Based on these
observations and arguments, the \textit{snake}-like path is
adopted in our practical simulations throughout the rest part of
the paper.

\begin{figure*}[!tbp]
\includegraphics[angle=0,width=1\linewidth]{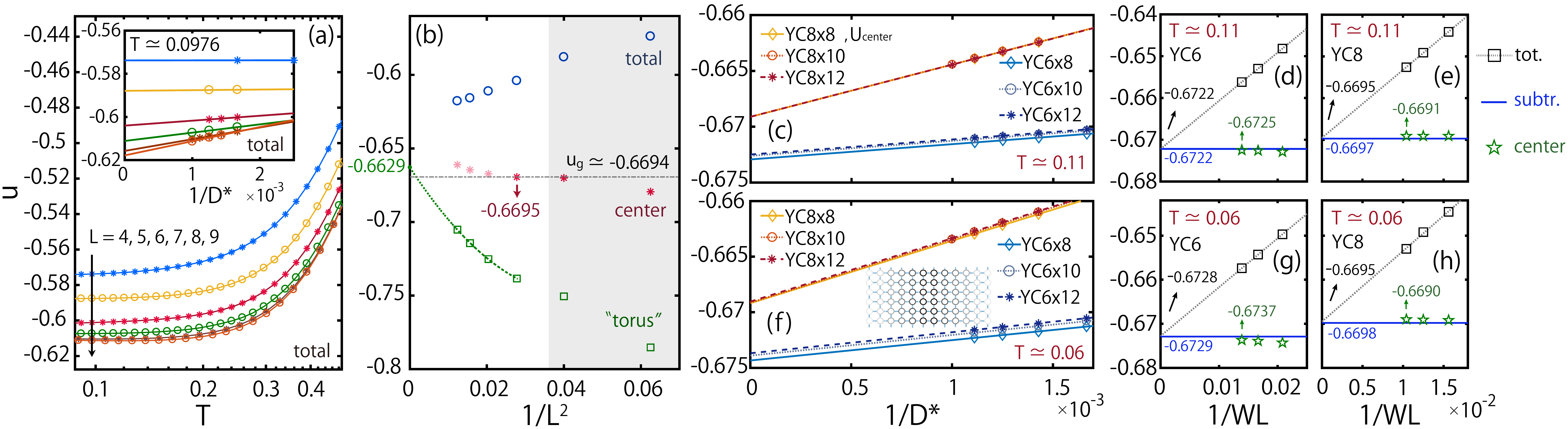}
\caption{(Color online)
   (a) Internal energy $u$ on the \OS{L}{L} lattices up to $L=9$
   calculated by XTRG keeping up to $D^*= 1000$ multiplets. The
   data are obtained in three different ways (total, central, and
   ``torus") as described in the main text, with {\it total} shown
   in main panel (a), and finite size scaling of the extrapolated
   data vs. $1/L^2$ shown in panel (b). In order to reduce the
   finite $D^*$ effects, we extrapolate the internal energy $u$ to
   $1/D^*=0$, as seen in the inset of (a). In (b) we collect the
   low-temperature ($T\simeq 0.1$) data extrapolated in (a)
   $1/\Dstar\to0$ and analyze it here vs. $1/L^2\to0$. For {\it
   ``torus"}, we extrapolate the four largest system sizes (i.e.,
   data in gray shaded area was excluded), to the thermodynamic
   limit, via a second-order polynomial fitting vs. $1/L^2$. The
   horizontal dashed line represents the ground state energy
   $u_g\simeq-0.6694$ from QMC \cite{Sandvik2010}.
   For comparison, (c-h) analyses the internal energy $u$ of SLH on
   \YC{6}{L} and \YC{8}{L} cylinders of lengths $L=8,10,12$.
   Exemplary extrapolations of $\ucenter$ vs. $1/\Dstar\to0$ are
   shown in panels (c) and (f) for $T\simeq 0.11$ and $0.06$,
   respectively, where $\ucenter$ is evaluated via a weighted
   average around the center as illustrated in the inset of (f)
   (see main text for more details). The results at $1/\Dstar\to0$
   are collected vs. $1/WL$ in panels (d, e, g, h) (green stars).
   There they are also compared to similarly extrapolated data for
   \utot (black squares), as well as to $u_{\rm{subtr}}$ (blue
   horizontal line) obtained by subtracting the length $L=8$ from
   the $L'=12$ cylinder. With \utot also extrapolated to $1/WL\to
   0$, we find good agreement across our data towards the
   thermodynamic limit.
}
\label{Fig:Cv-SF-OS}
\end{figure*}

\section{Square-Lattice Heisenberg Model}
\label{Sec:SLH}

In this and the next section we present our main thermodynamic
results for the SLH and the QIM, respectively. We benchmark them
against QMC data generated by the looper algorithm from ALPS
\cite{Bauer.b+:2011:ALPS}.

\subsection{Internal energy and specific heat}

In Fig.~\ref{Fig:W6uCv}, we present the results for the internal
energy $u$ as well as specific heat $c_V$, where we have
employed both XTRG and SETTN to simulate the SLH on two
lattices, \OS{6}{6} and \YC{6}{12}. We have also compared the
two schemes for computing $u$ and their derived $c_V$ in
Fig.~\ref{Fig:W6uCv}: (a) as expectation values by tracing the
Hamiltonian [cf. \Eq{Eq:uTr}], or (b) by taking the derivative
of free energy [cf. \Eq{Eq:uDr}].

The internal energy results $u$ obtained from both schemes agree
very well with the QMC data, as shown in Figs.~\ref{Fig:W6uCv}(a,c).
By strongly zooming in into the low-$T$ regime, nevertheless, it
turns out that scheme $b$ results in slightly better accuracy,
in both XTRG and SETTN simulations. Still given the same bond
dimension, within scheme $a$, XTRG data demonstrates better
accuracy than those of SETTN. This observation is consistent
with the general observation that XTRG produces more accurate
results due to the much smaller number of evolution and thus
truncation steps \cite{Chen2018} for the density matrix
$\rho(\beta/2)$.

The slight difference between the two schemes $a$ and $b$ is
arguably due to truncation: truncation is biased to keep the
strongest weights in $\rho$, such that
$\mathcal{Z}(\beta)=\mathrm{Tr}[\rho(\beta/2)^\dagger
\rho(\beta/2)]$ is optimally represented, hence also $f\sim -
\frac{1}{\beta} \ln \mathcal{Z}$, and thus also its derivative
$u$, i.e., as in scheme $b$. Conversely, by computing $u$
directly as in scheme $a$ via the expection value
$\mathrm{Tr}\bigl( \rho^\dagger H \rho \bigr)$, this is not
necessarily guaranteed to be optimally represented in the
presence of truncation. This heuristically explains the
slightly better performance of scheme $b$.

We also compare the specific heat $c_V$ derived from the
respective internal energy data obtained from both XTRG and
SETTN simulations in schemes $a$ and $b$. The results are shown
in Figs.~\ref{Fig:W6uCv}(b,d), with the same conclusion as for
the internal energy $u$: scheme $b$ leads to a slightly better
numerical performance for both RG methods. The peak position
for $c_V$ allows us to read off a characteristic crossover
temperature $T_s$ for the SLH, separating the low-temperature
regime showing incipient long-range order from a
high-temperature regime without such order (as discussed in
more details below).

To scale the results to the thermodynamic limit, we show the
internal energy $u$ of SLH on \OS{L}{L} lattices with $L= 4$ to
9 in \Fig{Fig:Cv-SF-OS}(a). We collect the energy values
calculated by scheme $b$ [Eq.~(\ref{Eq:uDr})] at our lowest
reliable temperature $T\simeq 0.1$, which already provides a
very good estimate of ground state energy \cite{Okabe1988}.
With the $u$ data well converged vs. $T$ on the finite-size
clusters, we extrapolate the energy results to $1/\Dstar\to0$ as
shown in the inset of \Fig{Fig:Cv-SF-OS}(a). Three slightly
different ways of extrapolating the ground state energy towards
the thermodynamic limit $1/L^2\to0$ are presented in
Fig.~\ref{Fig:Cv-SF-OS}(b): {\it total} (blue circles) is
obtained by dividing the total energy by the number of sites
$N=L^2$, {\it torus} (green squares) to be defined below, and {\it
center} (maroon asterisks). The latter is obtained from an
smooth average emphasizing center sites, computed as $\ucenter
\equiv \tfrac{1}{\sum_{i}{w_i}} \sum_{i}{w_i u_i}$, where $u_i$
is the energy per site which equals half the plain sum of
nearest-neighbor bond energies around the site $i$, and the
weighting factors are taken as $w_{i\equiv (i_x,i_y)} =
\sin^2\bigl(\frac{i_x-1}{L-1}\pi\bigr)
\sin^2\bigl(\frac{i_y-1}{L-1}\pi\bigr)$, with $i_x,i_y
\in[1,L]$. They are maximal in the center and smoothly diminish
towards the open boundary where they vanish quadratically, hence
suppressing the influence of the open boundary. This {\it
center} data converges fast vs. $1/L$. For $L=6$ it already
equals $-0.6695$ in excellent agreement with the QMC result
$u_g\simeq-0.6694$ (see, e.g., Ref.~\cite{Sandvik2010}).
However, for our largest system sizes, $L\gtrsim7$, the bond
energy distribution starts getting weakly affected by our
limited bond dimension $D^*$, e.g., see extrapolation in
$1/\Dstar$ in the inset of \Fig{Fig:Cv-SF-OS}(a). Thus \ucenter
starts to drift away from the plateau approximately reached for
$L\sim 6$ due to an increased error in the extrapolation
$1/\Dstar\to0$. A similar behavior is likely also observed for
the `total' data for the largest system sizes.

\begin{figure}[!tbp]
\includegraphics[angle=0,width=0.85\linewidth]{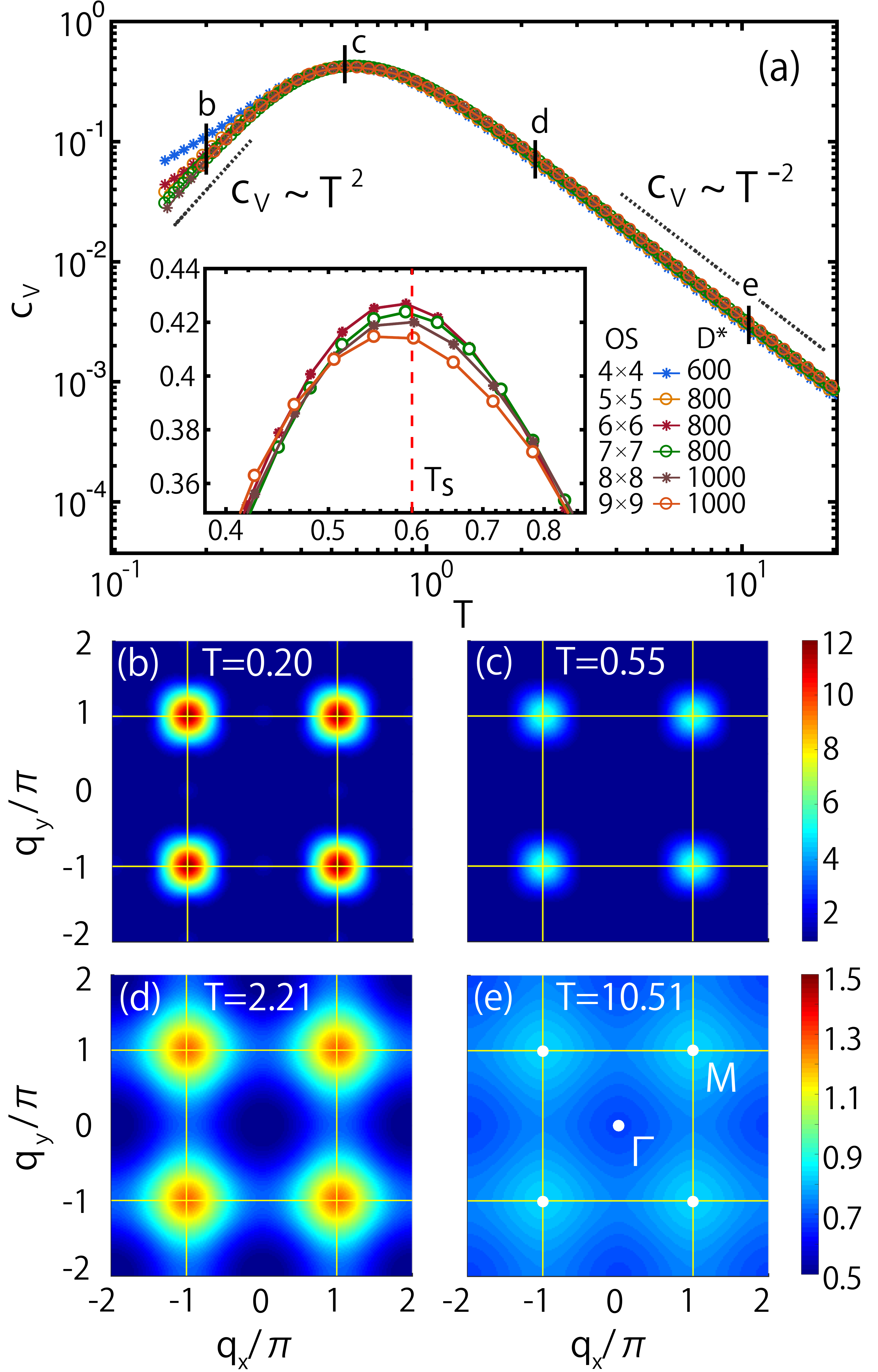}
\caption{(Color online) 
   (a) SLH specific heat $c_V$ on the \OS{L}{L} lattices on log-log
   scale to emphasize the algebraic behavior at high and low
   temperature, obtaining, $c_V \sim T^{-2}$ and $\sim T^{2}$,
   respectively (see dashed lines as guide to the eye). The inset
   zooms into the peak at intermediate temperatures around $\TS
   \sim 0.6$. Panels (b-e) show the static spin structure factors
   $S({q})$ on \OS{9}{9} at $T\simeq0.20$, $0.55$, $2.21$, and
   $10.51$, respectively. The grid lines demarcate the Brillouin
   zone, where the white dots in (e) indicate specific
   high-symmetry points therein. 
}
\label{Fig:u2}
\end{figure}

To further confirm the energy extrapolation, a fictitious {\it
torus} (green squares) is introduced, which also incorporates a
weighted average $\utorus = \tfrac{1}{N} \sum_{i} w'_i u_i$.
Here the weights $w'_i$, defined as
\[
   w'_{i}=\left\{ \begin{aligned}
      1\text{,} & \,\, &&{i \in \text{bulk}}\\
      {4}/{3}\text{,} & \,\, &&{i \in \text{edge}}\\
      2\text{,} & \,\, &&{i \in \text{corner }}
   \end{aligned}
\right.
\]
reflect the fact that boundary sites have missing bonds w.r.t. a
fictitious torus, i.e., a corner site has two bonds missing (so
we multiply the site energy by a factor of $w'_i=2$) and an edge
site one bond (thus $w'_i=4/3$). In a sense we are estimating
the energy values on a ``torus", by adding the missing bonds of
a given boundary site whose energies replicate existing
nearest-neighor bonds. This somewhat overestimates the energies
of the boundary sites, such that the ground state energy
converges from below now, as seen in \Fig{Fig:Cv-SF-OS}(b). We
extrapolate this data for the {\it ``torus"} only including the
data points of $L \geq 6$) to the thermodynamic limit via a
polynomial fitting. From this we obtain $u_g^*\simeq-0.6629$,
which is slightly above the QMC result.

For comparison, we also simulate YC geometries of widths $W{=}6$
and $8$ at two temperatures $T \simeq 0.11$ and $0.06$, with
their internal energy $u$ analyzed in \Figs{Fig:Cv-SF-OS}(c-h).
As seen in Figs.~\ref{Fig:Cv-SF-OS}(c,f) similar to the inset in
\Fig{Fig:Cv-SF-OS}(a), the convergence of \ucenter exhibits a
nearly linear behavior vs. $1/\Dstar$ and can thus be well
extrapolated to $1/\Dstar=0$.
The extrapolation over $1/D^*$ may also be replaced by an
extrapolation of the truncation error, i.e., the discarded
weight $\delta\rho \to 0$. We show in \App{App:TruncErr} for
the case of YC6 and YC8 at $T \simeq 0.06$ that both
extrapolations agree well at low temperatures.

Similar to \Fig{Fig:Cv-SF-OS}(b) we compare the internal energy
in three different ways in \Figs{Fig:Cv-SF-OS}(d,e,g,h) (again
all extrapolated to $1/\Dstar\to0$), except that the earlier
fictitious {\it torus} is replaced by a {\it subtracted} data
set for $u_{\rm{subtr}}$ (horizontal lines) which is obtained
from the difference in \utot between 
$L=8$ and $L'=12$
cylinders, divided by $(L'-L) \cdot W$ sites. 
Also for the case of cylinders, \ucenter is the energy per site
weighted by a factor $\sin^2(\frac{i_x-1}{L-1}\pi)$ that is
uniform around the cylinder, i.e., independent of $i_y$, with
$i_x \in [1,L]$ indexing columns along the cylinder. The
weights are illustrated in the inset of
Fig.~\ref{Fig:Cv-SF-OS}(f), where the intensity gradually
decreases from the center to both ends. Besides this ``smooth"
average, we have also tried the computation of $\utot$ sharply
restricted within the 1$\sim$2 central columns of the cylinder,
yielding slightly less systematic results. 
Overall, the results of all three schemes are in good agreement
with each other, as well as with the QMC data $u_g\simeq-0.6694$
\cite{Sandvik2010}. For example, in the case of YC8 at $T\simeq
0.06$, $\utot = -0.6695$, $\usubtr = -0.6698$, and $\ucenter =
-0.6690$, leading to an accurate estimate of ground state energy
$u_g^*=-0.6694(4)$.

The derivative of the internal energy yields the specific heat
$c_V$ [cf. \Eq{Eq:Cv}], shown for the SLH in \Fig{Fig:u2}(a) on
\OS{L}{L} lattices up to $L=9$. 
We observe a well-pronounced
single peak located at $\TS \sim 0.6$. Given that is largely
independent of system size (see inset), this data already
reflects the thermodynamic limit (even though simulating finite
system sizes!). This observation is consistent with the
scenario that there is no phase transition in SLH at finite $T$
and, consequently, that $\TS$ represents a crossover scale of
thermodynamic behavior.

\subsection{Static Structure Factor}

Next, we explore the spin structure factor $S({q})$ at various
temperatures. We select four temperatures corresponding to
different regimes in the specific heat [see markers (b-e) in
\Fig{Fig:u2}(a)], and show their $S({q})$ data in
Figs.~\ref{Fig:u2}(b-e), respectively. High symmetry points in
the Brillouin zone (BZ) including the central point
$\Gamma=(0,0)$ and $M=(\pm\pi,\pm\pi)$ are indicated explicitly
in \Fig{Fig:u2}(e). At low temperature $T\ll \TS$, there exists
a clearly established incipient order, which gives rise to the
sharp bright spots at the $M$ points.
As $T$ increases, the system passes the cross-over scale $T\sim
\TS$, at which stage the incipient order has already become
significantly weakened, as shown in Fig.~\ref{Fig:u2}(c).
As the temperature increases further, the originally bright spot
at the $M$ points becomes ever weaker [\Figs{Fig:u2}(d,e),
note also the altered color bar scale], until it is completely
blurred out for temperatures $T>10$ [\Fig{Fig:u2}(e)].

\begin{figure}[!tbp]
\includegraphics[angle=0,width=1\linewidth]{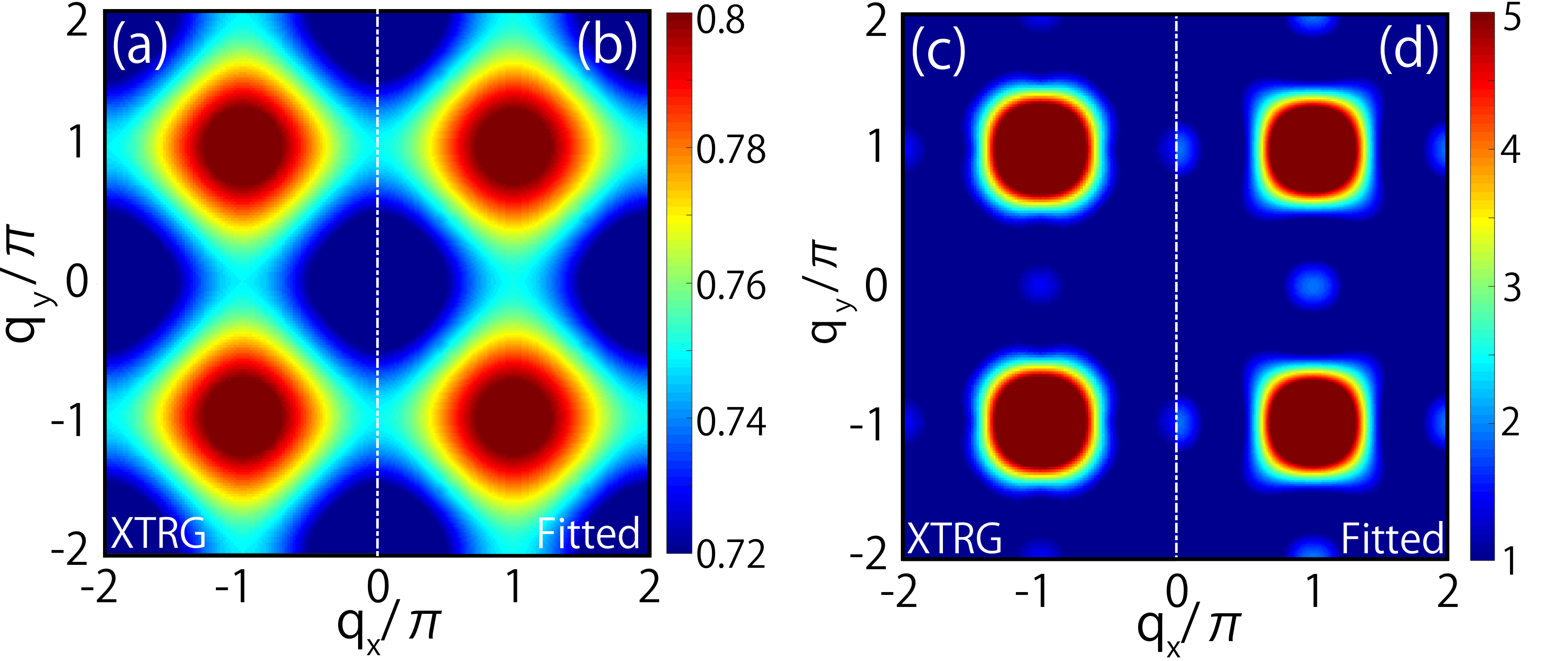}
\caption{(Color online)
   The static spin structure factors $S({q})$ on the \OS{9}{9}
   lattice at (a) $T \simeq 12.5$ and (c) $T \simeq 0.2$,
   calculated by XTRG. Fittings using the antiferromagnetic Ising
   (AFI) model and independent dimer approximation (IDA) (see main
   text) are presented at the right half of each panel [(b) and
   (d)], which enjoy excellent agreement with XTRG data on the left
   half of each panel [(a) and (c), respectively].
}
\label{Fig:SiandSd}
\end{figure}

Besides the bright $M$ points, the $S(q)$ contour shows
nontrivial patterns near the crossover scale. We illustrate
this on the example of an \OS{9}{9} SLH system in \Fig{Fig:SiandSd}.
It zooms in the low-intensity part of $S(q)$, showing salient
patterns in stark difference between the high- and
low-temperature regimes. At high temperature $T \simeq 12.5$
[\Figs{Fig:SiandSd}(a,b)], there exists a clear-edged ``diamond"
shape surrounding the $M$ points. On the other hand, in the
low-temperature regime, e.g. $T \simeq 0.2$
[\Figs{Fig:SiandSd}(c,d)], the diamonds have significantly shrunk
and rotated by $45^\circ$.

In order to get a better intuitive understanding, we employ two
simple models, the independent dimer approximation (IDA) and the
antiferromagnetic Ising (AFI) model. In IDA, we assume that a
given site is in a singlet configuration with either one of its
nearest-neighbor sites with probability $1/4$ for each, and no
further longer-range correlations. This yields the spin
structure factor \begin{equation} S_D(q_x, q_y) =
\tfrac{3}{8}\left(2-\cos q_x-\cos q_y\right), \label{Eq:Sd}
\end{equation} which describes short-range correlation
(typically at high $T$). On the other hand, the AFI spin
structure factor $S_I$ is evaluated from spin correlations of
classically ordered antiferromagnet configurations on an
\OS{5}{5} lattice, to capture the essential feature in the
spin-spin correlation at low temperatures. 

Indeed, at high temperatures we find that a fit of the form
$S(q) = a S_D + c$ based on IDA with parameters $a=0.08$ and
$c=0.69$ [as shown in \Fig{Fig:SiandSd}(b)] provides a good
description of the XTRG data in Fig.~\ref{Fig:SiandSd}(a). From
this we conclude that at high temperatures $T\gg \TS$, IDA can
reproduce the diamond pattern and capture very well the residual
magnetic correlations in the system. Note that, by definition,
the $q$-independent term in $S(q)$ must be equal to
$\mathbf{S}^2=3/4$, hence $\tfrac{3}{4} a+ c \simeq 0.75$ [cf.
\Eq{Eq:Sd}] with $a\ll 1$ at large $T$.

At low temperatures, we employ the AFI correlation
introduced above to describe the developed incipient order, 
together with the dimer correlations taking care of
the short-range fluctuations, again under IDA assumption. 
The structure factor is therefore then fitted using
the combination $S({q}) = a S_{D} + b S_{I}+c$, 
where we find that $a=4.5$, $b=25$, and $c= -2.7$ well resembles
the XTRG $S({q})$ data (larger values of $a$ and $b$ suggest
longer-range correlations as expected, indeed), including even
the very subtle details of the four-leaf shape.

For pure long-range AFI correlations, the $q_x$ and $q_y$
components decouple in the structure factor into a product of
independent terms, such that $S(q)$ develops square-like peaks
around the $M$ points that are aligned with the BZ. At high
temperatures, instead, the lines are aligned with the smaller
magnetic BZ boundary, given that the real-space lattice unit
cell is enlarged. This explains why the diamond pattern in
\Figs{Fig:SiandSd}(a-b) rotates into aligned square like peaks in
\Figs{Fig:SiandSd}(c-d). In this sense, the inclusion of $S_D$,
i.e., short-range correlations, is important to allow four
little ``leaves" to appear (which may disappear in the
thermodynamic limit, though). We believe, however, that the
dominant features seen in the $S(q)$ contours in \Fig{Fig:SiandSd},
indeed, encode important information on the spin structures in
the system. Apart from the different brightness of $S(M)$, this
feature constitutes another relevant distinction in $S(q)$
between the high- and low-temperature regimes. We expect that
these salient patterns in $S(q)$ may find their experimental
realizations in quantum simulators using cold atoms
\cite{Mazurenko2017}.

\begin{figure}[!tbp]

\includegraphics[angle=0,width=1\linewidth]{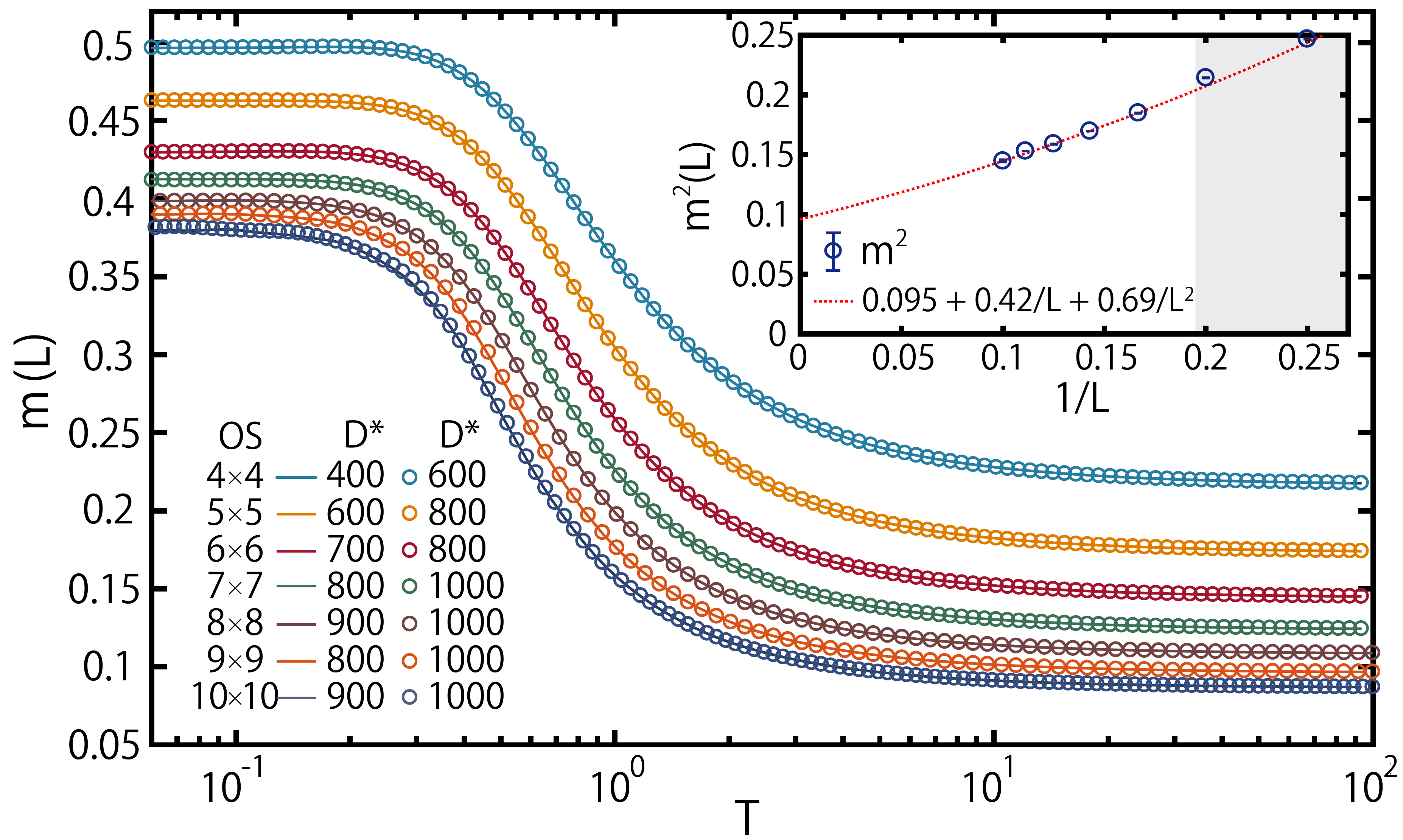}
\caption{(Color online)
   The finite-size analysis of spontaneous magnetization
   $m(L)\equiv\sqrt{S(M)/L^2}$ of the SLH on the \OS{L}{L} lattices
   up to $L=10$. 
   In the inset, we collect $m^2(L)$ data at
   $T=0.1$ where convergence vs. $T$ is reached, and extrapolate it
   to the thermodynamic limit $1/L\to0$ using a parabolic fit. The
   data in the gray shaded area was excluded from this fit. From
   this we estimate the value for the thermodynamic limit $\mS^*
   \simeq 0.30(1)$, in good agreement with the literature,
   $\mS\simeq0.3070(2)$ \cite{Sandvik-1997}.
}
\label{Fig:MagOS}
\end{figure}

\subsection{Spontaneous Magnetization and Incipient Order at $T>0$}

From the spin structure factor $S({q})$ at the ordering point
$q_0\equiv M$, we can estimate the spontaneous magnetization
$\mS$. To be specific, we employ the low-temperature
finite-size spontaneous magnetization $m(L) = \sqrt{S(M)/L^2}$
on the \OS{L}{L} as an estimate, which is shown as a function of
$T$ in \Fig{Fig:MagOS}. We find for all systems explored,
including our largest system at $L=10$, that $m(L)$ has
essentially saturated at low temperatures $T\lesssim0.1$. We
then collect the converged values of $m(L)$, plot $m^2(L)$ vs.
$1/L$ \cite{Sandvik-1997} in the inset of \Fig{Fig:MagOS}. A
quadratic fit in $1/L$ then yields the estimate $\mS^* \simeq
0.30(1)$. This XTRG result is in good agreement with the QMC
value $\mS\simeq0.3070(2)$ \cite{Sandvik-1997}.

\begin{figure}[!tbp]
\includegraphics[angle=0,width=0.9\linewidth]{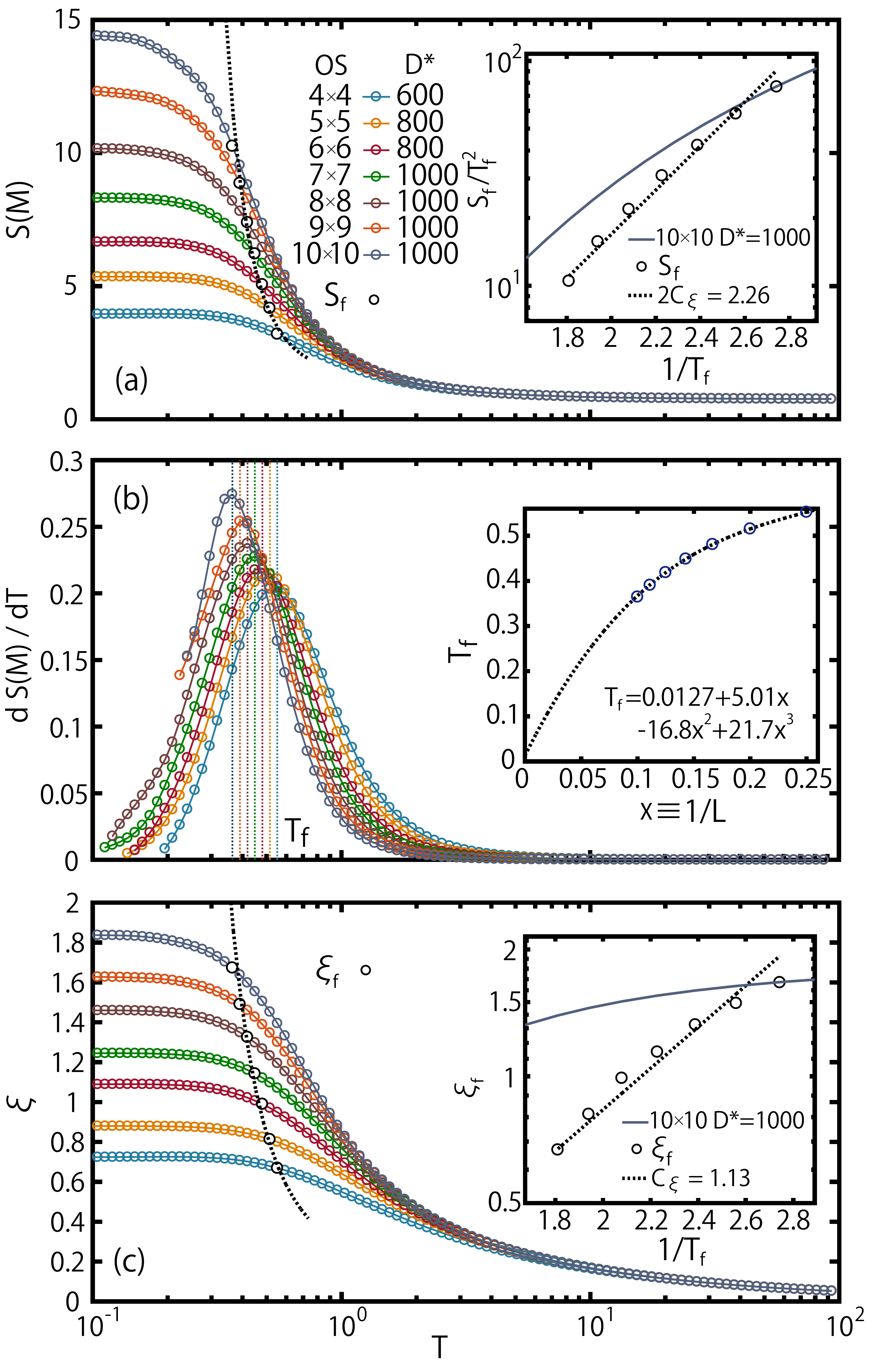}
\caption{(Color online)
   (a) Static structure factor $S(M)$ vs. $T$ (same data as in
   \Fig{Fig:MagOS}). Its derivative in (b) has a maximum which
   defines a specific temperature $T_f$ at which {\it finite size}
   effects become significant. The inset in (a) then analyzes
   $S_f/T_f^2$ vs. $1/T_f$. 
   The inset in (b) shows that $T_f$ extrapolates to $0$ in the
   thermodynamic limit, via a cubic fit as shown based on the data
   with $L\geq4$.
   (c) Correlation length $\xi$ vs. $T$. The results for $S(M)$
   and $\xi$ evaluated at $T_f$, are marked by black circles in the
   main panels (a,c), respectively. The data is collected and
   analyzed in the respective insets vs. $1/T_f$ which, overall,
   shows good agreement with RC predictions (dashed line). For
   comparison, the insets in (a,c) also plot the data for S(M) and
   $\xi$ vs. $1/T$ for the largest system size.
}
\label{Fig:RC}
\end{figure}

\subsection{Renormalized Classical Behaviors}

At low temperatures, $T\ll \TS$, the SLH enters the universal RC
regime \cite{Chakravarty1988,Chakravarty1989}. As observed in
large-scale QMC simulations \cite{Beard1998,Kim1998}, as well as
in neutron scattering experiments \cite{Chakravarty1988}, the
incipient order and RC scalings have been quantitatively
confirmed. To be specific, the correlation length $\xi$
diverges exponentially with decreasing $T$ as
\begin{equation}
   \xi(T) = A_{\xi}\, e^{C_{\xi}/T} [1+ \mathcal{O}(T)],
\label{Eq:Xiexp}
\end{equation}
and the structure factor $S({q})$ also diverges at the ordering
point as
\begin{equation}
   S(M) = A_{S}\, T^2 e^{2C_\xi/T} [1+ \mathcal{O}(T)], 
\label{Eq:Sexp}
\end{equation}
where $A_{\xi}$, $A_{S}$, and $C_{\xi}=2\pi \rho_s$ are
constants, with $\rho_s$ the spin stiffness \cite{Elstner1994}.

The universal RC scalings in Eqs.~(\ref{Eq:Xiexp},
\ref{Eq:Sexp}) are strictly valid only in the thermodynamic
limit $L \to \infty$. In our \OS{L}{L} XTRG simulations, we
only have finite-size thermal data up to $L=10$, such that below
(some) low temperature our finite-size XTRG data necessarily
will deviate from the exponential scalings. This occurs once
the thermal correlation length reaches system size. It coincides
with the temperature where the structure factor $S(M)$ starts to
saturate which was already clearly observed in \Fig{Fig:MagOS}
[replotted in \Fig{Fig:RC}(a) directly as $S(M)$ itself].

We may also use this as a criterion to define the (maximal)
thermal correlation length that fits into a given finite system.
Based on this then, we may analyze the onset of RC behavior from
our finite-size data. We start by estimating a temperature
$T_f$ below which the finite (f) size effects become prominent.
We define it as the temperature at which the derivative
$dS(M)/dT$ shows a maximum, as indicated by the vertical dashed
lines in \Fig{Fig:RC}(b). Being due to finite size effects, a
polynomial fitting vs. $1/L$, as shown in the inset of
\Fig{Fig:RC}(b), shows that $T_f\to0$ for $1/L\to0$, as expected.
Next we collect $S(M)$ evaluated at $T_f$ (denoted as $S_f$)
from various \OS{L}{L} systems. A semilog plot, shown in the
inset of Fig.~\ref{Fig:RC}(a), shows that this approximately
supports an exponentially diverging behavior, indeed. This
notably differs from the data for $S(M)_T$ simply plotted vs.
$1/T$ for the largest system size, also shown for comparison
(blue line). While for large temperatures (smaller $1/T$) the
slope on the log-plot approximately coincides with the earlier
$S_f$ analysis, it shows clear deviations due to finite size at
lower temperatures (larger $1/T$). This is in contrast to the
analysis vs. $1/T_f$ which was designed to largely eliminate
finite size effects. We have compared the $S_f$ vs. $T_f$ curve
to the standard RC formula Eq.~(\ref{Eq:Sexp}) with
$C_{\xi}=2\pi \rho_s\simeq1.13$ \cite{Elstner1994}, as indicated
by the dashed lines in both the main panel and the inset of
Fig.~\ref{Fig:RC}(a). The remarkable agreement strongly
suggests that the RC behavior can be uncovered in the
finite-size data via a careful analysis.

Similar to the analysis of the structure factor $S_f$ at the $M$
point, one can compute the (maximal) correlation length $\xi_f
\equiv \xi(T_f)$ as shown in \Fig{Fig:RC}(c). The resulting RC
behavior of $\xi_f$ vs. $1/T_f$ is shown in the inset. In the
present case, it is well-fitted by \Eq{Eq:Xiexp}, thus again
supporting RC scaling.

\begin{figure}[!tbp]
\includegraphics[angle=0,width=0.85\linewidth]{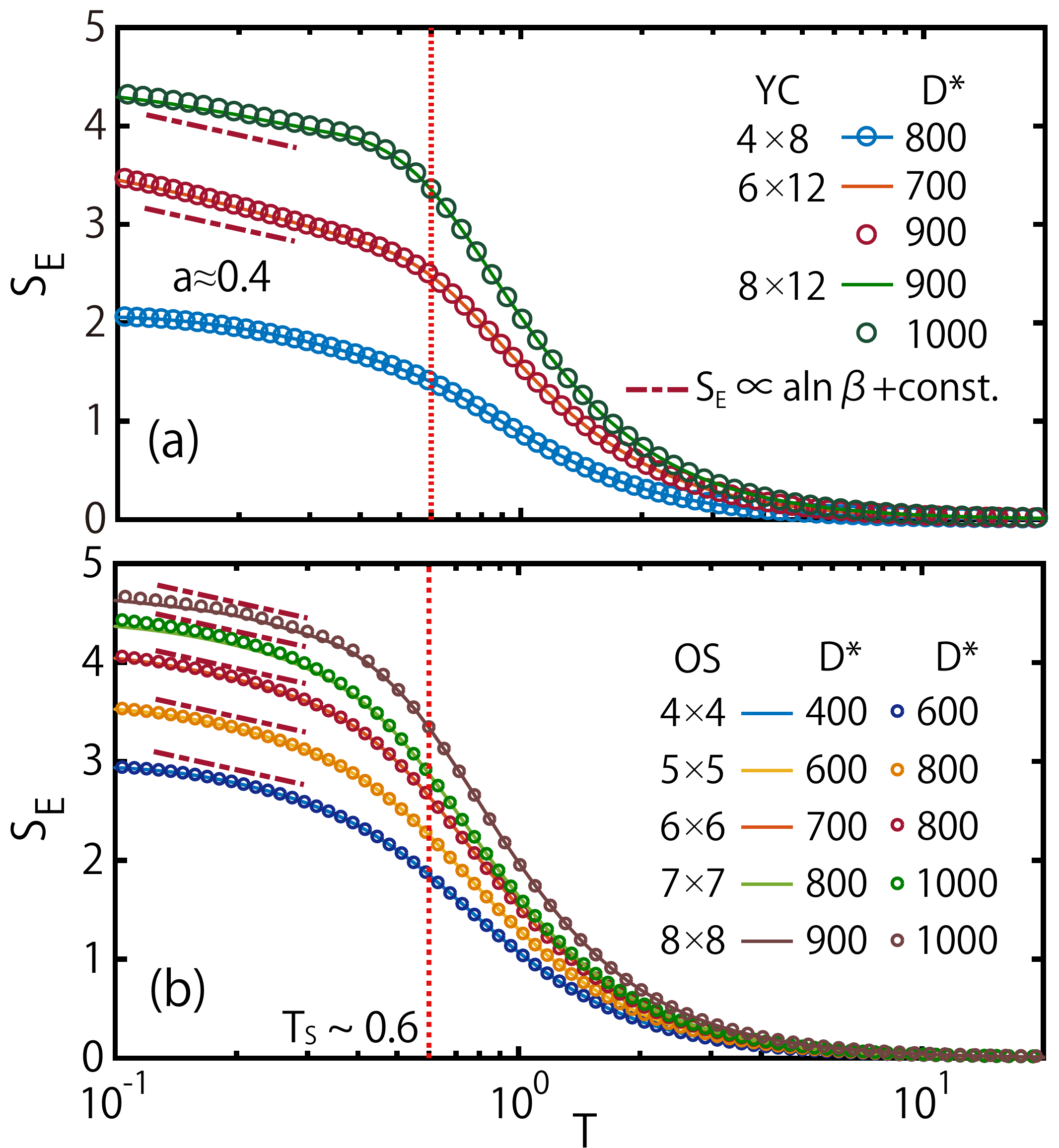}
\caption{(Color online)
   MPO entanglement entropy $S_E$ on (a) cylinders \YC{4}{8},
   \YC{6}{12}, and \YC{8}{12}, and (b) \OS{L}{L} with $L=4$ to $8$.
   The tilted dashed lines in both (a) and (b) represent $S_E
   \propto a \ln{\beta} +\mathrm{const.}$, with slope $a \simeq
   0.4$ (see dashed guides on top of the curves). The vertical
   dotted line represents the temperature scale $\TS \sim 0.6$
   [e.g. see \Fig{Fig:u2}(a)].
}
\label{Fig:SLHSE}
\end{figure}

\subsection{Entanglement Scaling}

Low-temperature logarithmic scalings in the entanglement entropy
$S_E$ have been observed in a number of quantum systems with
gapless excitations. Near the conformal critical points in 1D
quantum systems, the entanglement entropy scales like $S_E = a
\ln{\beta} + \mathrm{const.}$, with $a$ proportional to the
conformal central charge \cite{Chen2018,Chen2018a,
Barthel.t:2017:FiniteT,Dubail17,Prosen.t+:2007:Entropy,
Marko.z+:2008:Complexity}. For 2D quantum systems with gapless
Goldstone modes, e.g., the triangular lattice Heisenberg
antiferromagnet \cite{Chen2018a}, logarithmic entanglement also
appears and can be related to a tower of states due to the
ground-state SU(2) symmetry breaking. On intuitive grounds, one
may expect a slowdown of the entanglement entropy at low
temperatures, bearing in mind that the classical AF ground state
is a product state.

In Fig.~\ref{Fig:SLHSE}, we plot the thermal entanglement
entropy $S_E$ vs. $T$, for YC and OS geometries. In
Fig.~\ref{Fig:SLHSE}(a), despite a rapid (algebraic) decrease at
high temperatures, $S_E$ ``crosses over" into a logarithmic
behavior in the low-temperature regime around $T<\TS$, with an
estimated slope of $a \simeq 0.4$ approximately independent of
the system width (note that, in contrast, the temperature
independent offset in $S_E$ {\it is} roughly proportional to the
system width). The transition temperature is consistent with
the crossover scale $\TS\sim0.6$ that had been identified from
the peak position in the specific heat, e.g., see
\Fig{Fig:W6uCv} or \Fig{Fig:u2}(a).
Hence, from \Fig{Fig:SLHSE} we find that the incipient AF order
for $\TS\sim0.6$ is directly linked to a weak logarithmic
scaling of the entanglement entropy vs. $T$. For the OS
systems in \Fig{Fig:SLHSE}(b) we find stronger finite-size
effects with an onset of saturation at our smallest
temperatures, qualitatively similar to what is already also
visible for our smallest \OS{4}{4}. Still also for the OS
systems, we find approximately the same logarithmic scaling of
$S_E$ with the same slope as for the cylinders in
\Fig{Fig:SLHSE}(a) for $T<\TS$.

\section{Magnetic Phase Transition in the Quantum Ising Model}
\label{App:QIM}

In this section we study the QIM as an exemplary minimal model
system that exhibits a finite temperature phase transition. It
thus constitutes a very meaningful benchmark for XTRG. While
not explicitly analyzed here, at $T=0$, the square-lattice QIM
also possesses a QPT at a critical field $h_c=1.52219(1)$,
between the paramagnetic and ferromagnetic phases
\cite{TFI2DQMC,CTMRG-2009, CTMRG-2012}. Finite-temperature
properties of the QIM have also been explored by TPO simulations
\cite{Czarnik.p+:2012:PEPS, Czarnik.p+:2015:PEPS} in the
thermodynamic limit. 

\begin{figure}[!tbp]
\includegraphics[angle=0,width=0.95\linewidth]{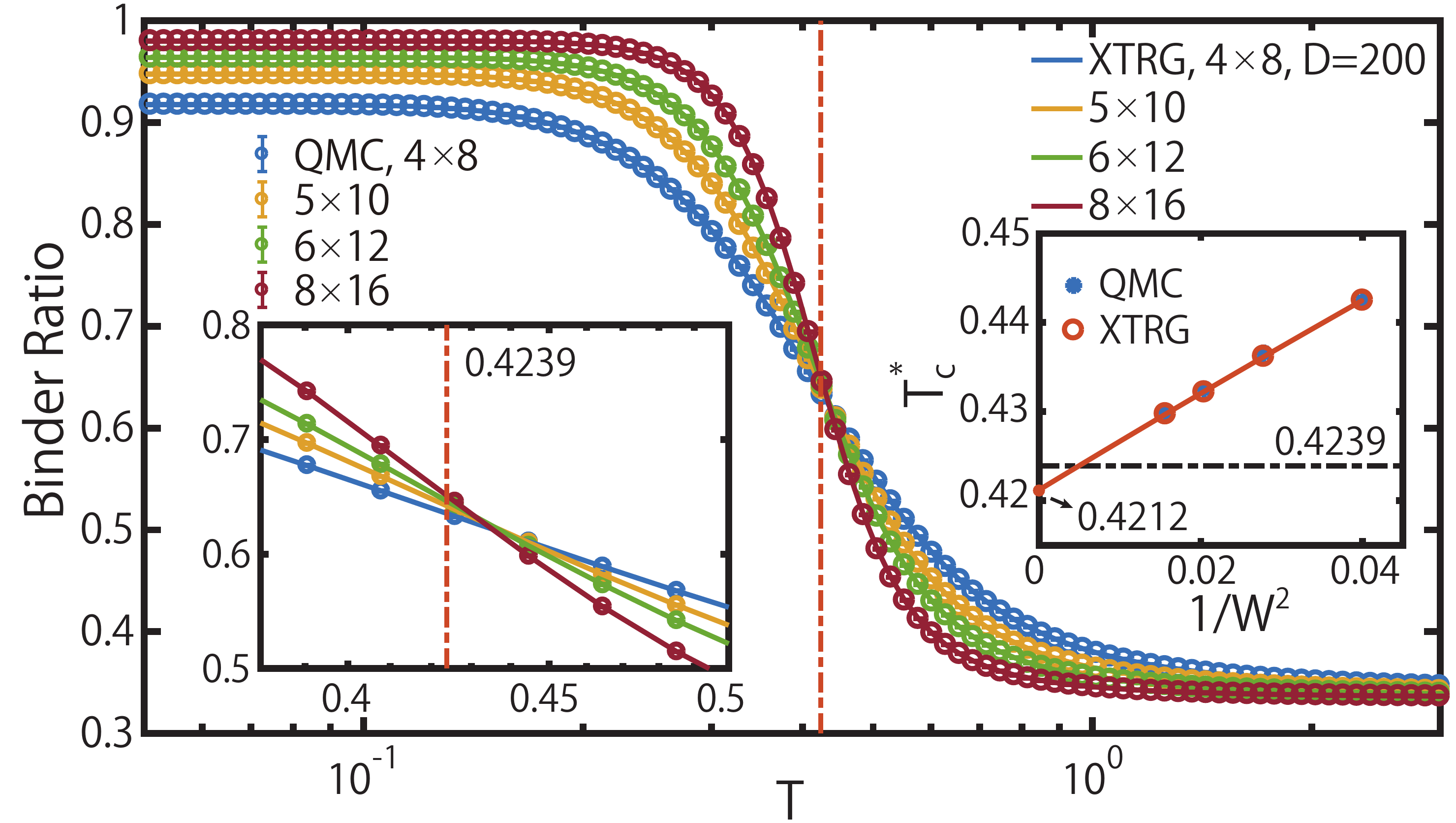}
\caption{(Color online)
   Binder ratio curves of QIM for the same $h_x = \frac{2}{3} h_c$
   for \YC{W}{L} systems with a fixed aspect ratio $L/W=2$. Left
   inset zooms in the region near the crossing points $T_c^*$, and
   right inset shows a subsequent second-order polynomial
   extrapolation of $T_c^*$ to $1/W^2\to0$ (data $W=7$, not shown
   in the main panel for better readability, is included for the
   right inset). A second-order polynomial fitting as shown
   yields an extrapolated $T_c^*(0) \simeq0.4212$, which is in
   excellent agreement with $T_c \simeq 0.4239$ in the
   thermodynamic limit \cite{Czarnik.p+:2015:PEPS}.
} 
\label{Fig:qIsingBR}
\end{figure}

We show XTRG results for the QIM [Eq.~(\ref{Eq:QIM})] in
\Figs{Fig:qIsingCv}--\ref{Fig:qIsingEntLand} for YC geometries
up to width $W=8$ with a fixed aspect ratio $L/W=2$, as well as
\OS{L}{L} with $L$ up to 10. Due to the transverse field, the
system only possesses $\mathbb{Z}_2$ symmetry. We focus on the
fixed value $h_x=\frac{2}{3} h_c$ of the transverse field where
the model exhibits a thermal transition at the critical
temperature $T_c\simeq0.4239$ \cite{Czarnik.p+:2015:PEPS}.
There we analyze various thermal quantities of interest,
including the specific heat $c_V$, Binder ratio $U_4$, and the
MPO entanglement $S_E$. We exploit them to study the
finite-temperature phase transition. A detailed comparison to
QMC is performed, whenever the latter is available. 

\subsection{Specific heat}

\begin{figure}[!tbp]
\includegraphics[angle=0,width=0.95\linewidth]{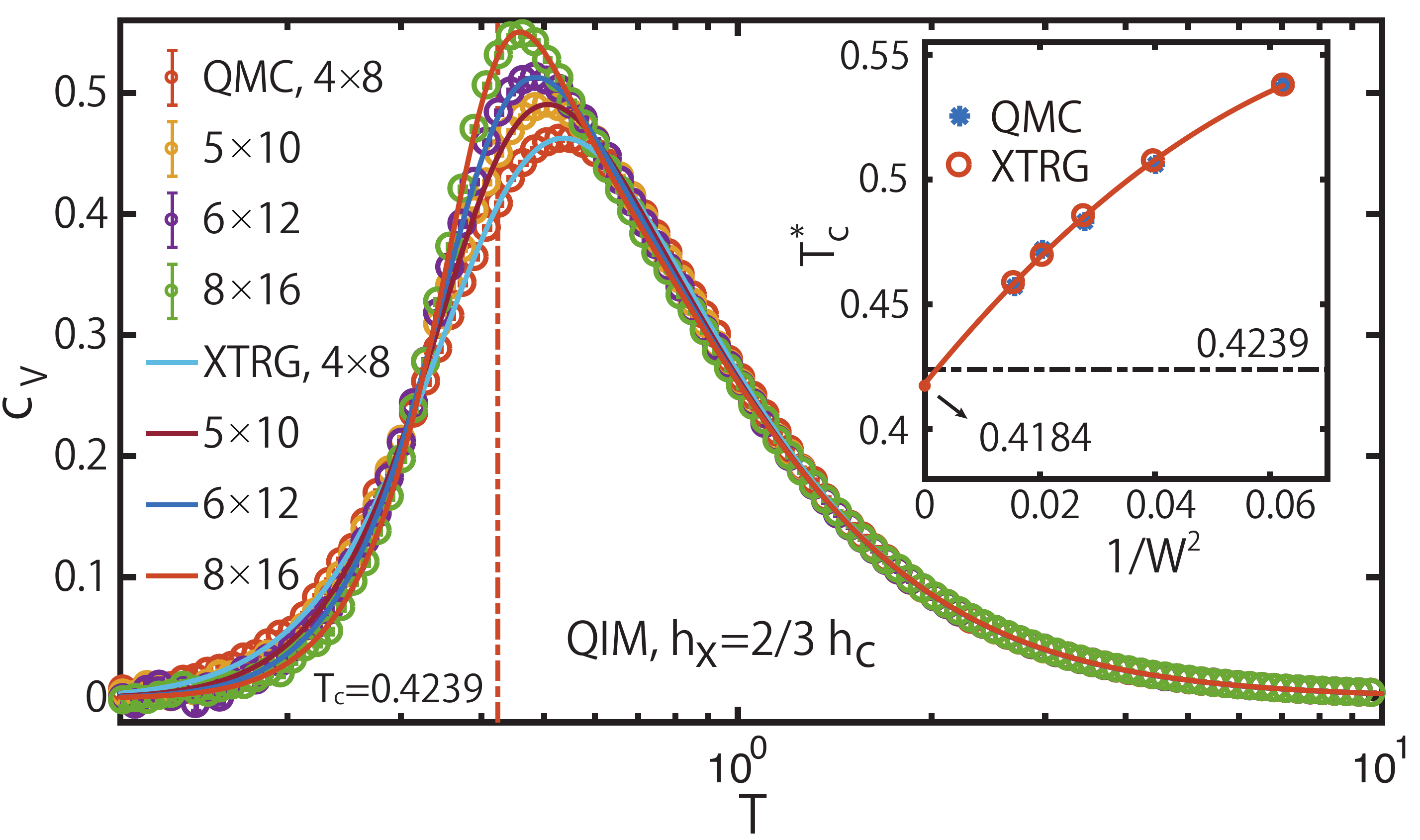}
\caption{(Color online)
   Specific heat of QIM for fixed $h_x = \frac{2}{3} h_c$ on
   \YC{W}{L} of width up to $W=8$ and length $L=2W$ ($W=7$ curve
   not shown in the main panel for better readability). The XTRG
   results retaining up to $D=240$ states coincide with the QMC
   reference data. In the inset, we collect the peak position
   $T_c^*$ of $c_V$ curves calculated by QMC and XTRG, which also
   coincide, and extrapolate towards the exact critical temperature
   in the thermodynamic limit $x\equiv 1/W^2\to0$ by a second order
   polynomial fit, having $T_c^*(x) \simeq -14.5\, x^2 + 2.8\, x +
   T_c^*(0)$, with an extrapolated value of $T_c^*(0)\simeq0.4184$
   (a polynomial fit in $1/W$ leads to a similar value).
}
\label{Fig:qIsingCv}
\end{figure}

The specific heat $c_V$ from XTRG is compared to standard QMC
data on YC geometries up to $W=8$ in \Fig{Fig:qIsingCv} with
excellent overall agreement. Note that for XTRG we only
retained a moderate number of at least $D=200$ bond states to
reach convergence. Due to the thermal phase transition, the
specific heat for finite-size systems shows a single-peak, the
height of which becomes more and more pronounced as $W$
increases. We track the position $T_c^*$ of this peak, and
analyze it in the inset vs. $1/W^2\to0$. The data for $T_c^*$
from both methods virtually coincides, thus supporting the
quality of the data. For the thermodynamic limit $1/W^2\to0$ we
obtain $T_c^*(0) \simeq 0.4184$ which differs by about 1.3\%
from the value $T_c\simeq0.4239$ obtained in
\cite{Czarnik.p+:2015:PEPS}.

\subsection{Binder ratio and phase transition temperature} 

However, extracting the thermal transition temperature from
plain thermal quantities such as peak position int the specific
heat in the previous section, gives rise to larger finite-size
errors. According to the finite-size scaling (FSS) theory,
higher moments, such as Binder cumulants, offer a more accurate
means for determining $T_c(0)$. One widely adopted Binder
cumulant in QMC simulations is
\begin{equation}
   U_4 = \frac{ \langle (S_{\mathrm{tot}}^z)^2 \rangle^2}
              {\langle (S_{\mathrm{tot}}^z)^4 \rangle},
\label{Eq:U4}
\end{equation}
where $S_{\mathrm{tot}}^z = \sum_i S_i^z$ is the total spin.
The Binder ratio $U_4$ has significantly smaller finite-size
corrections, namely, $\sim L^{-2}$
\cite{Binder-1981a,Binder-1981b}. 
To be specific, according to Ginzburg-Landau theory, the total
magnetization of block spins, i.e., $M_z = \sum_i \langle S_i^z
\rangle_{\beta}$, obeys the Gaussian distribution. In the
infinite $T$ limit, it is easy to verify, via Gaussian
integration, that $U_4=1/3$, while for the $T \rightarrow 0$
limit, it trivially tends to $U_4=1$. Right at $T_c$, according
to the FSS theory, $U_4$ flows to a nontrivial fixed value,
i.e., it stays as a constant as the system size $N$ increases
(given $N$ large). Therefore, $U_4$ curves for different system
sizes cross at $T_c^*$, providing a very accurate determination
of the critical temperature $T_c$.

With MPO techniques, the two expectation values and their ratio
$U_4$ in Eq.~(\ref{Eq:U4}) can be obtained very conveniently.
The total moment operator $S_{\mathrm{tot}}^z$ has a simple MPO
representation of bond dimension $D=2$, from which one can
construct an exact representation of $(S_{\mathrm{tot}}^z)^2$
(with $D=3$) and $(S_{\mathrm{tot}}^z)^4$ ($D=5$) at ease.

In Fig.~\ref{Fig:qIsingBR} we show the calculated Binder ratio
by XTRG and QMC, which again show excellent agreement in both
the main panel and insets. The left (bottom) inset zooms in the
region in the vicinity of the cross point. Taking the crossing
temperature $T_c^*(W)$ of two curves $W$ and $W+1$ as an
estimate of the critical temperature, two $T_c^*$ data sets are
extracted from QMC and XTRG, and plotted vs $1/W^2$ in the right
inset. Again XTRG and QMC data are virtually on top of each
other. The estimate from our largest system size results in
$T_c^*(W=7) \simeq 0.4297$. A second-order polynomial
extrapolation $1/W^2\to0$ yields $T_c^* \simeq 0.4212$, which
agrees with the thermodynamic limit in
\cite{Czarnik.p+:2015:PEPS} to within 0.6\%.

\subsection{Thermal Entanglement}

\begin{figure}[!tbp]
\includegraphics[angle=0,width=0.95\linewidth]{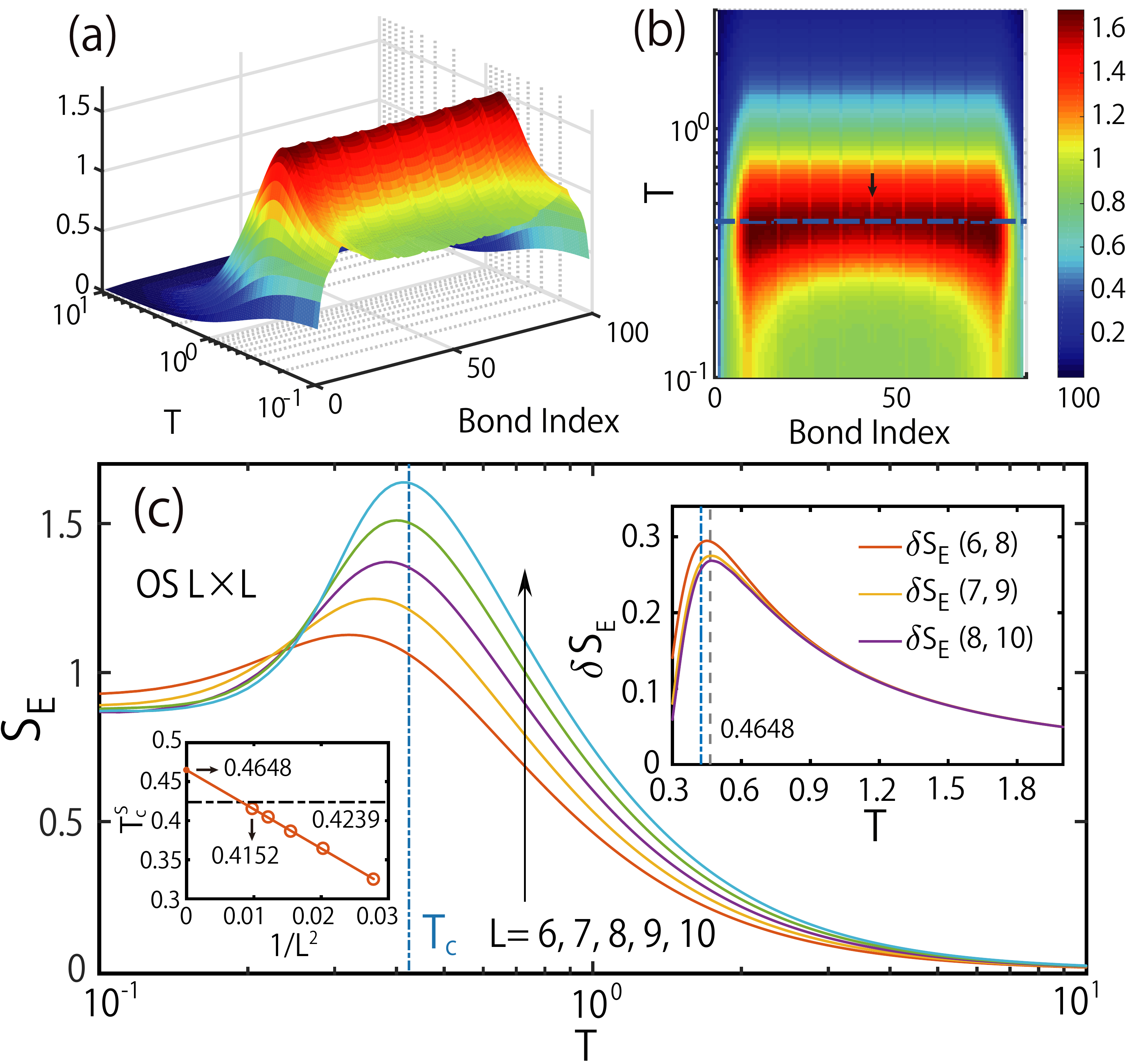}
\caption{ (Color online)
(a) Entanglement landscape of QIM thermal states on the
   \OS{10}{10} lattice, for the same $h_x = \frac{2}{3} h_c$, vs.
   bond indices and temperature.
(b) Top view, showing that the peak
   temperatures at central bonds 
   (away from boundaries) locate right at
   the critical temperature $T_c$ (horizontal dashed line).
(c) $S_E$ vs. $T$ on various \OS{L}{L} lattices, cut at a
   central bond [indicated by the the black arrow in (b)] with
   maximal $S_E$. The blue vertical dashed lines
   in both the main panel and right inset indicate $T_c \simeq
   0.4239$ in the thermodynamic limit \cite{Czarnik.p+:2015:PEPS}.
   The left inset shows the peak temperature $T_c^S$ 
   vs. inverse system size $1/L^2$, 
   which approach the true critical temperature
   $T_c$. $D=240$ bond states are kept in the calculations.
   The right inset shows the entanglement difference vs. $T$
   between consecutive system sizes i.e.,  
   \OS{L}{L} and \OS{L'}{L'} with $L'=L+2$ bearing in mind
   even-odd effects.
   The gray vertical dashed line corresponds to the first-order
   extrapolated value at $1/L^2\to0$ in the left inset.
}
\label{Fig:qIsingEntLand}
\end{figure}

In the QIM case with a thermal phase transition towards a gapped
low-temperature phase, the entanglement entropy features a
maximum around the transition temperature. Here we also examine
the scaling of MPO entanglement $S_E$ vs. $T$ for different
bonds at which the system is cut when computing $S_E$. The
resulting ``entanglement landscape" is shown in
\Fig{Fig:qIsingEntLand}(a) where we observe a clear ridge line
along $T \simeq T_c$, i.e., the surmised peak in $S_E$ at $T_c^S
\simeq T_c$. The shape and location of this peak appears
stationary in the center of the system (modulo width of the
system), yet varies slightly towards to open boundaries
[Fig.~\ref{Fig:qIsingEntLand}(b)]. This suggests that the peak
position $T_c^S$ in the bulk can be taken as a good estimate of
critical temperature $T_c$ of the thermal phase transition. 

In Fig.~\ref{Fig:qIsingEntLand}(c) therefore we show slices of
the entanglement landscape for the bond in the center of the
system that maximizes $S_E$. Now as we increase $L$ the peak
becomes more and more pronounced, and the finite-size estimate
of critical temperature $T^S_c$ approaches the critical
temperature $T_c$. However, as seen from the inset of
Fig.~\ref{Fig:qIsingEntLand}(c), while the finite-size $T_c^S$
appears well-suited for extrapolation in $1/L^2\to0$ in
principle, when doing so, the resulting value based on the
present data would actually significantly overshoot the true
critical temperature, as $T_c^S \simeq 0.4648$.
A similar behavior is also seen on YC
geometries (not shown). Hence, so far $S_E$ does not lend itself
to an simple extrapolation to obtain an accurate critical
thermal transition temperature.

In order to gain some insight into the systematic overshooting 
in the extrapolation of $T_c^S$, 
we plot in the right inset of Fig.~\ref{Fig:qIsingEntLand}(c)
the entanglement difference $\delta S_E(L,L') = S_E(L') - S_E(L)$ 
between \OS{L'}{L'} and \OS{L}{L} lattices, 
with $L'$ and $L$ both even or odd (to avoid even-odd oscillation). 
There are a number of features important for analyzing $S_E$. 
The lines in the inset lie on top of each other for $T>1$, 
meaning $dS/dW \sim \mathrm{const}$ there
in agreement with an area law for the entanglement entropy.
Moreover, given that the difference $\delta S_E$ for the smallest
system sizes in our data upper bounds $\delta S_E$ for larger
systems, 
$S_E/W$ does not 
diverge at $T_c$, but stays finite, which is in stark contrast, e.g.,
to the specific heat data. 

Moreover, from the analysis in the inset
the peak position in the $\delta S_E$ 
data appears to remain above $T_c$
in the thermodynamic limit, 
already consistent with
the extrapolated $T_c^S$ 
for $1/L^2 \to 0$ in
the left inset of Fig.~\ref{Fig:qIsingEntLand}(c). 
Much of this behavior appears related
to the strong asymmetry in $S_E$ 
due to a gapped low-temperature phase.
Therefore, for
the accurate determination of $T_c$ from $S_E$, 
it appears one needs to come up with a different procedure
other than just extrapolating the temperature for the
maximum in $S_E$.
Nevertheless, it is an interesting observation that from an
entanglement point of view,
the maximum in $S_E$ can systematically occur above $T_c$
even in the thermodynamic limit.
The precise location may depend on the geometry, i.e.,
boundary conditions and aspect ratio of the system,
and as such deserves further studies.

\section{Summary}

In this work, we have employed two TTN algorithms, the
SETTN and XTRG approaches, to investigate two prototypical
quantum spin models, the square-lattice Heisenberg and
transverse-field Ising models. We explore four conventional MPO
paths, finding that the \textit{snake}-like path constitutes an
overall favorable choice, due to its smaller entanglement and
thus less truncation errors on long cylinders and stripes.

Throughout, we found excellent agreement of SETTN and XTRG data
with QMC results of both models. Based on these accurate
finite-size thermal data of SLH, we are able to extrapolate to
the groundstate energy $u_g^*\simeq -0.6694(4)$ (from YC8
results), as well as the spontaneous magnetization $\mS^* \simeq
0.30(1)$, all of which are in good agreement with large-scale
QMC results. We extract the well-established renormalized
classical behaviors, i.e., the exponential divergence at low
$T$, of the structure factor $S(q)$ and correlation length $\xi$
vs. $T$, at the ordering momentum $M$.

We have also explored the thermal entanglement $S_E$ in the MPO
representations of the equilibrium density matrices. $S_E$
exhibits a logarithmic scaling in the SLH, which is
likely related to gapless excitations in the model. For QIM with a
finite-$T$ phase transition, $S_E$ shows a pronounced peak at
$T_c^S$, where the thermal phase transition takes place. 
Besides, $T_c^*$ from the crosspoint of Binder ratio
curves provides very accurate estimate, i.e., down to
below 1\% errors, of the critical temperature $T_c$ in the
thermodynamic limit.

Our benchmark calculations reveal that TTN methods, such
as XTRG, are highly efficient and accurate in solving quantum
many-body problems at finite $T$. Besides the unfrustrated SLH
and QIM systems explored in detail here, XTRG can be applied to
more challenging frustrated quantum magnets
\cite{Chen2018,Chen2018a}. There it may play an essential role
in bridging the gap between experimental thermal data of
currently numerous spin liquid candidate materials and their
microscopic spin models.

\begin{acknowledgments}

This work was supported by the National Natural Science
Foundation of China (Grant No. 11834014) and Deutsche
Forschungsgemeinschaft (DFG, German Research Foundation) under
Germany's Excellence Strategy\,-\,EXC-2111\,-\, 390814868.
WL and HL are indebted to Qiao-Yi Li for stimulating discussions.
BBC was supported by the German Research foundation, DFG
WE4819/3-1. AW was funded by DOE DE-SC0012704. 
\\
\end{acknowledgments}

\appendix

\setcounter{equation}{0}
\setcounter{figure}{0}
\setcounter{table}{0}

\makeatletter

\renewcommand{\theequation}{A\arabic{equation}}
\renewcommand{\thefigure}{A\arabic{figure}}
\renewcommand{\theHfigure}{A\arabic{figure}}
\renewcommand{\bibnumfmt}[1]{[#1]}
\renewcommand{\citenumfont}[1]{#1}

\begin{figure*}[htb]
\centering
\includegraphics[width=0.9\linewidth]{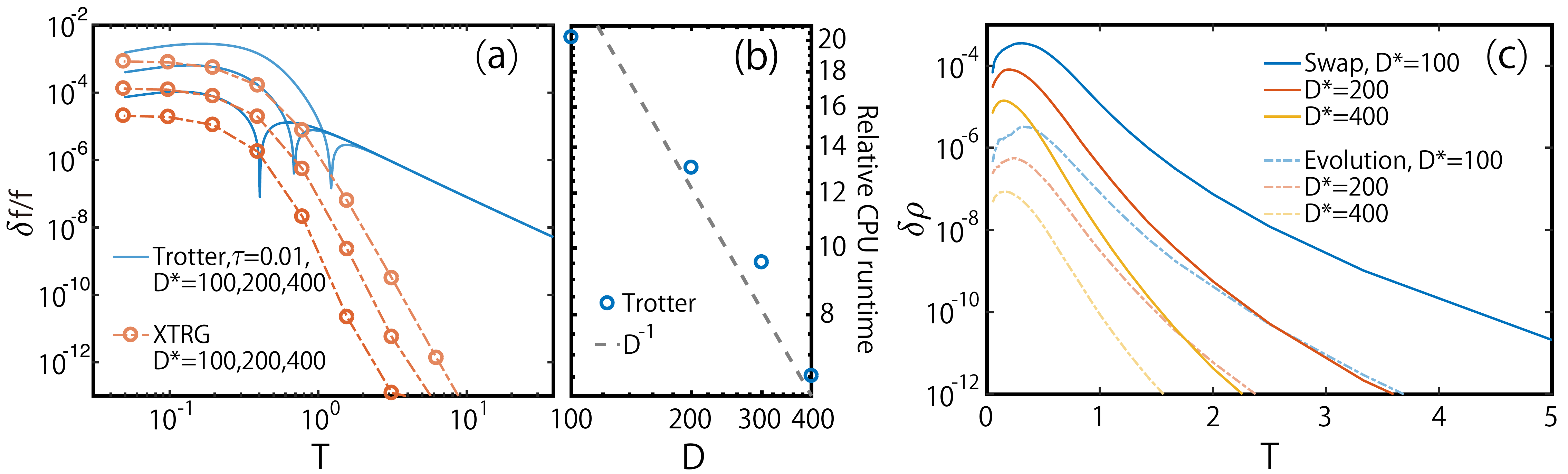}
\caption{(Color online)
(a) Comparison of XTRG and Trotter linear
   evolution (with swap gates) schemes in the benchmark calculations of
   an SLH model on \OS{4}{4}. 
(b) Computational runtime of the Trotter approach
   relative to XTRG, which scales roughly as $1/D$
   (with the dashed line a guide to the eye).
(c) The truncation error analysis in the
   linear Trotter calculations: the swap gates contribute significantly
   larger (over two orders of magnitude) truncation errors than those
   of imaginary-time evolution gates.
}
\label{Fig:LinVSExp}
\end{figure*}

\begin{figure*}[htb]
\includegraphics[angle=0,width=0.95\linewidth]{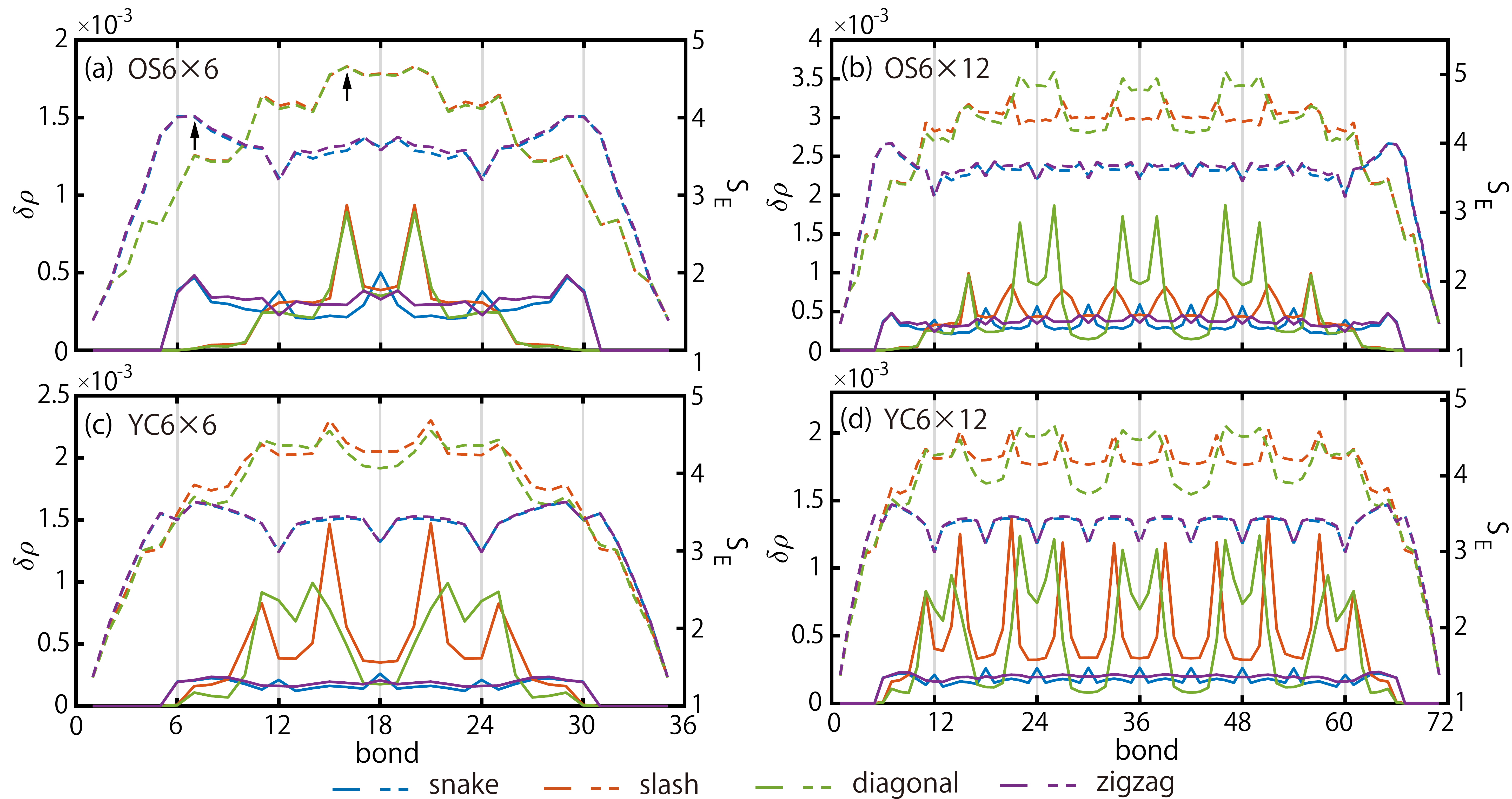}
\caption{(Color online)
   Truncation error (solid lines; left axis set) and entanglement
   entropy $S_E$ (dashed lines; right axis set) at a low
   temperature ($T\simeq0.06$). The data are obtained by retaining
   $\Dstar=500$ multiplets. We compared the four different mapping
   paths in \Fig{Fig:maxPe}, namely, the \textit{snake}-like, 
   \textit{zigzag}, \textit{diagonal}, and \textit{slash}
   on (a) \OS{6}{6} (b) \OS{6}{12} (c) \YC{6}{6} and (d)
   \YC{6}{12}. The entanglement entropy is precisely also the data
   that was used for the width of the lines in \Fig{Fig:maxPe} in a
   graphically more organized way.
}
\label{Fig:truncPe}
\end{figure*}

\section{Exponential tensor renormalization group vs. Trotter and swap gates}
\label{Sec:Compare}

In thermal tensor network simulations, we start from infinite
temperature, $\beta=0$, where $\rho(0)$ has a trivial
representation as direct product of identity matrices, to
various lower-temperature mixed states. The most straightforward
way is to perform such a linear imaginary-time evolution using
Trotter gates, which has been widely used \cite{Li.w+:2011:LTRG,
Ran.s+:2012:Super-orthogonalization,Dong.y+:2017:BiLTRG,
Czarnik.p+:2012:PEPS,Czarnik.p+:2015:PEPS,Czarnik.p+:2016:TNR,
Czarnik.p+:2017:Sign}. When applied to 2D systems, given an MPO
representation of the density matrix, additional auxiliary swap
gates have to be introduced, as adopted in 2D thermal RG methods
based on minimally entangled typical thermal states 
\cite{Bruognolo.b+:2017:MPS}. 

On the other hand, a very efficient scheme, XTRG, was proposed
in Ref.~\cite{Chen2018}, where we cooled down the system
exponentially following Eq.~(\ref{Eq:XTRG}). In
Fig.~\ref{Fig:LinVSExp}, we compare XTRG to the linear evolution
with Trotter and swap gates on the SLH model
[Eq.~(\ref{Eq:SLH})] on an \OS{4}{4} geometry. In
Fig.~\ref{Fig:LinVSExp}, we choose $\tau=0.01$ in the Trotter
decomposition, which constitutes a good compromise in terms of
Trotter error relative to truncation error and overall runtime.

As shown in Fig.~\ref{Fig:LinVSExp}(a), XTRG is found to be more
accurate compared to the Trotter scheme, given the same bond
dimension. For example, Fig.~\ref{Fig:LinVSExp}(a) shows that
the Trotter data with $D^*=400$ ($D\simeq1600$) yield similar
accuracy as XTRG with $D^*=200$ ($D\simeq800$). In addition, the
(relative) CPU hours are plotted in Fig.~\ref{Fig:LinVSExp}(b),
showing that the Trotter scheme is slower than XTRG by roughly
one order of magnitude. As seen on the log-log scale,
however, the relative Trotter performance improves with
increasing $D$ roughly as $1/D$, in agreement with the
fact that the Trotter approach scales like $O(D^3)$ and
whereas XTRG as $O(D^4)$.
In order to exploit the reduced truncation error
with increasing $D$, though, Trotter would also have
to reduce the Trotter error by decreasing the 
Trotter time step, which likely offsets some of the
apparent gain with increasing $D$ (note that XTRG
is free of Trotter error). Specifically, also note
that there is a sign change in $\delta f$ for Trotter,
as seen by the downward kink in the $\log |\delta f|$
plot in \Fig{Fig:LinVSExp}(a), which moves towards
lower temperatures with increasing $D$. Having $\delta f$
change its sign is an indication that the
Trotter error is dominant down to lower temperatures,
before truncation error sets in.

We explicitly also analyzed truncation and swap gate errors in
the 2D Trotter approach in \Fig{Fig:LinVSExp}(c). The
truncation error due to swap gates (which help bring together
two spins with ``long-range" interactions after 1D mapping) are
about two orders of magnitude greater than those produced
directly in the imaginary-time evolution. Therefore, from
Fig.~\ref{Fig:LinVSExp}(c) we observe that the Trotter approach
in 2D accumulates significant swap-gate truncation error, and
thus it is not competitive in both efficiency and accuracy.

\section{Entanglement Entropy and Truncation Errors in Various MPO Paths}
\label{App:MPOPath}

Here we provide more detailed information on the entanglement
and truncation errors on each MPO bond. In \Fig{Fig:truncPe}, we
show them on four lattices including the \OS{6}{(6,12)} and
\YC{6}{(6,12)}, where the same $S_E$ data was also used in
\Fig{Fig:maxPe} to visually demonstrated the entanglement along
the various mapping paths. The present discussion therefore
extends the analysis of the mapping paths in \Sec{Sec:Geo}.

Quite generally, in \Fig{Fig:truncPe} the truncation error
$\delta\rho$ is largest where the block entanglement $S_E$ is
largest, such that peaks coincide 
(particularly for the \textit{slash} and \textit{diagonal} paths). 
On the \OS{6}{6} and
\YC{6}{6} lattices, the \textit{slash} and \textit{diagonal}
paths show peaks in the central part while the \textit{zigzag}
and \textit{snake}-like lines peak near both ends [indicated by
arrows in Fig.~\ref{Fig:truncPe}(a)]. Note, however, the
\YC{6}{6} case is already seen to be different from that on the
\OS{6}{6}. In Fig.~\ref{Fig:truncPe}(c), the \textit{slash} and
\textit{diagonal} lines have larger entanglement as well as
truncation errors, than those in the \textit{zigzag} and
\textit{snake}-like paths, not only in the very center but also
extended to regions near both ends.

For lattices with larger length $L$, the entanglement and
truncation peaks appear periodically in the bulk for all
mappings. As illustrated in Figs.~\ref{Fig:truncPe}(b,d), the
\textit{zigzag} and \textit{snake}-like paths show peaks still
near the open boundary and behave rather uniformly in the bulk.
This is in contrast to the \textit{slash} and \textit{diagonal}
paths which have higher $S_E$ overall, and thus perform
(considerably) worse.

\section{Data extrapolation vs. truncation error $\delta \rho$}
\label{App:TruncErr} 

\begin{figure}[tb!]
\includegraphics[angle=0,width=1\linewidth]{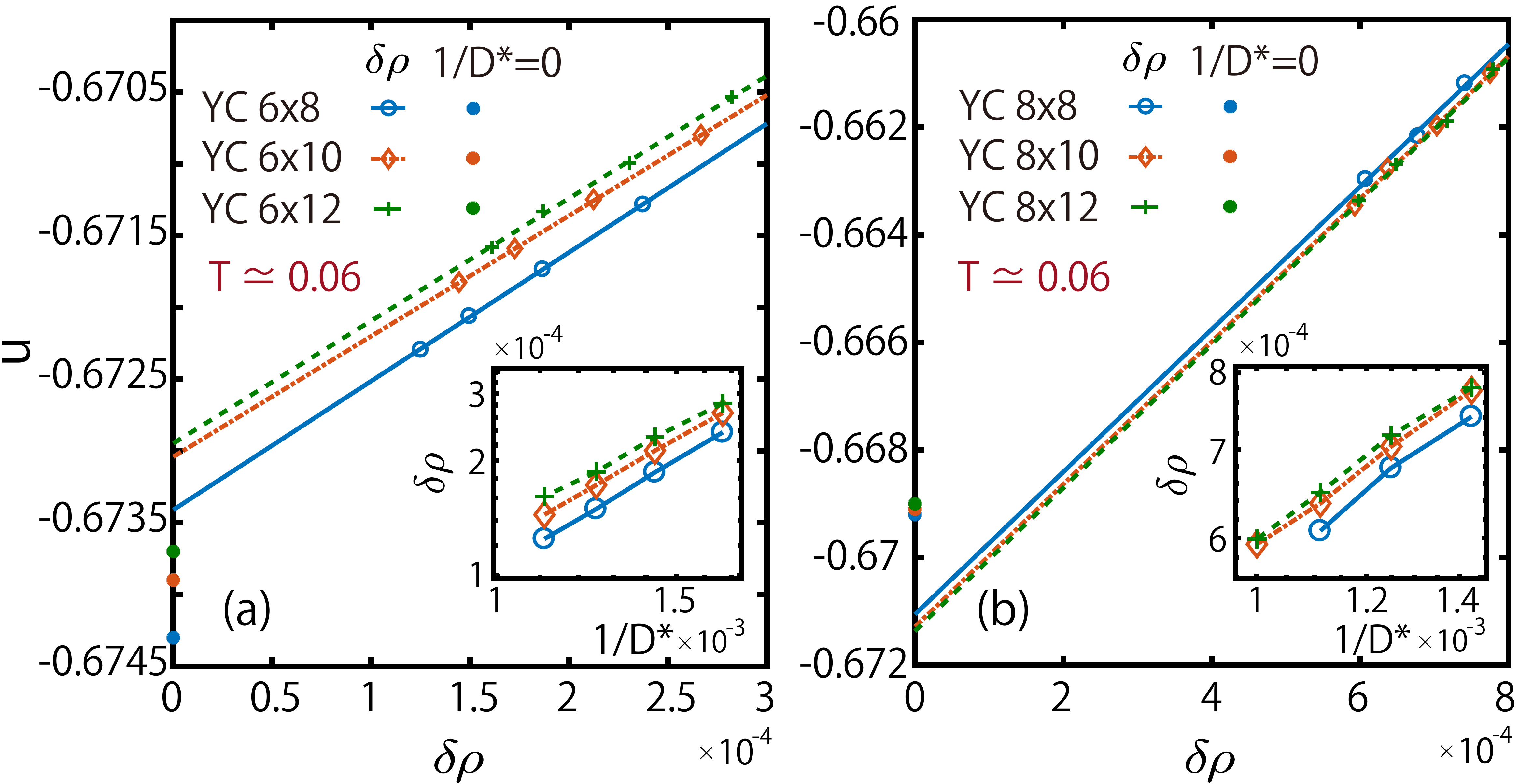}
\caption{(Color online)
   The energy per site $u$ vs. truncation error $\delta \rho$ on
   (a) YC6 and (b) YC8 geometries at $T\simeq 0.06$ [same as in
   \Fig{Fig:Cv-SF-OS}(f) of the main text]. A clear linear $u$ vs.
   $\delta \rho$ relation is observed and employed to perform the
   extrapolation. The thus extrapolated $u$ values are in a very
   good consistency to \mbox{$1/\Dstar{\to}0$} analysis
   (asterisks), with relative error $\sim$ 0.1\%. Insets in (a,b)
   show $\delta \rho$ vs $1/D^*$ on a log-log scale, showing
   polynomial scaling. 
}
\label{Fig:uvsdrho}
\end{figure}

In XTRG simulations, we can only retain a finite number of
multiplets $D^*$. This introduces a truncation error
$\delta \rho$ in the MPO representation of the many-body density
matrix. We showed in Figs.~\ref{Fig:Cv-SF-OS}(c,f) that
the low-temperature results for our largest cylinders (say, YC6
or 8) are no longer fully converged, in that
for example the internal energy $u$ still varies by
about 1\% when extrapolating $1/\Dstar\to0$

Nevertheless, to get a flavor on how reliable the extrapolation
vs. $1/\Dstar\to0$ is, we do a similar analysis here, but vs.
$\delta \rho\to0$, which represents the truncation error across
the geometric bond in the middle of the MPO. In
Fig.~\ref{Fig:uvsdrho}, we show $u$ vs. $\delta \rho$ for the
YC6 and YC8 lattice of various lengths, at fixed temperature
$T/J\simeq 0.06$.

Having sufficiently large \Dstar (sufficiently small
$\delta\rho$), similar to the $1/\Dstar$ extrapolation in the
main paper, we find an approximate linear relationship between
$u$ and $\delta \rho$ which can be extrapolated to
$\delta\rho\to0$, equivalent to the infinite $D^*$ limit.
The results are compared to extrapolated data in $1/$\Dstar 
in Figs.~\ref{Fig:Cv-SF-OS}(c, f) of the main text, where a good
agreement can be seen, for either YC6 and YC8 cases.

The linear relation $\delta u \propto \delta\rho$ can be
understood as follows. The truncation error $\delta \rho$ in
density matrix $\rho(\frac{\beta}{2})$ directly translates into
an error of the partition function $\mathcal{Z}(\beta) =
\rm{Tr} \bigl[\rho^\dagger(\frac{\beta}{2}) \,
\rho(\frac{\beta}{2})\bigr]$, since the latter precisely
resembles the cost function itself for optimizing
$\rho(\frac{\beta}{2})$. This argument is hand-wavy, of course,
since to be specific, we choose for $\delta \rho$ the truncation
error after a two-site variational optimization of MPO in the
center of the system. This is a good estimate for the accuracy,
but does not necessarily represent the precise full error in the
calculation of $\mathcal{Z}(\beta)$. Following thermodynamic
relations, finally, $\delta \rho$ also reflects linearly in the
errors of free energy and energy values, i.e., we can argue that
also $\delta f$, and therefore $\delta u \propto \delta \rho$
for small $\delta \rho$. 

In practice, for more challenging cases, due to the reason that
$\delta \rho$ only serves as an approximate estimate of
truncation not fully representing the errors in the variational
optimization, we find the analysis of $u$ vs. $1/D^*$
numerically more stable and accurate, which is therefore adopted
in Fig.~\ref{Fig:Cv-SF-OS} of the main text.

\bibliography{2DSquare}

\end{document}